# Generative AI Enhances Team Performance and Reduces Need for Traditional Teams

Ning Li (Tsinghua University)
lining@sem.tsinghua.edu.cn

Huaikang Zhou (Tsinghua University)
zhouhk@sem.tsinghua.edu.cn

Kris Mikel-Hong (Tsinghua University)
hongtb20@mails.tsinghua.edu.cn

Notes: We express our sincere gratitude to our research assistants Ming Jin and Meiling Huang for their invaluable contribution to running the laboratory. We acknowledge the technical support provided by Ziyu Deng, Yizhou Ji in analyzing text data. We also thank Junyuan Liu, Wenming Deng, Ziyan Cui, Jiatan Chen for their assistance in data and information collection. Our appreciation extends to all the participants who took part in the experiments. This study was supported by Tsinghua University School of Economics and Management Research Grant (No. 2022051001) and Tsinghua University Research Grant (No. 2023THZWJC01).

**Abstract**

Recent advancements in generative artificial intelligence (AI) have transformed collaborative work processes, yet the impact on team performance remains underexplored. Here we examine the role of generative AI in enhancing or replacing traditional team dynamics using a randomized controlled experiment with 435 participants across 122 teams. We show that teams augmented with generative AI significantly outperformed those relying solely on human collaboration across various performance measures. Interestingly, teams with multiple AIs did not exhibit further gains, indicating diminishing returns with increased AI integration. Our analysis suggests that centralized AI usage by a few team members is more effective than distributed engagement. Additionally, individual-AI pairs matched the performance of conventional teams, suggesting a reduced need for traditional team structures in some contexts. However, despite this capability, individual-AI pairs still fell short of the performance levels achieved by AI-assisted teams. These findings underscore that while generative AI can replace some traditional team functions, more comprehensively integrating AI within team structures provides superior benefits, enhancing overall effectiveness beyond individual efforts.

1. **Introduction**

Throughout history, teams have been the cornerstone of human achievement, enabling us to tackle complex tasks that require diverse ideas and collaborative efforts (Woolley, Chabris, Pentland, Hashmi, & Malone, 2010; Wu, Wang, & Evans, 2019; Wuchty, Jones, & Uzzi, 2007). With the rise of generative artificial intelligence (AI), a transformative shift is underway, where AI entities are not just tools but active participants in team processes (Hagemann, Rieth, Suresh, & Kirchner, 2023; McNeese, Schelble, Canonico, & Demir, 2021; Seeber et al., 2020). While the positive impact of generative AI on individual performance is well-documented (Dell'Acqua et al., 2023b; Jia, Luo, Fang, & Liao, 2023; Noy & Zhang, 2023), its scalability within a team setting remains poorly understood. To harness the power of generative AI and leverage human collaboration, we need clarity on how AI effects aggregate at scale within teams.

Existing research has explored the implications of technology on team performance, focusing primarily on human-AI interactions in specific functional contexts (Dubey, Hardy, Griffiths, & Bhui, 2024; Hosny, Parmar, Quackenbush, Schwartz, & Aerts, 2018). Prior to the invention of generative AI, the conventional focus in team dynamics research was largely on human-centric strategies, where technology served as an incremental enhancer rather than a core component (Grønsund & Aanestad, 2020; Puranam, 2021). Technologies were typically considered as supplementary aids that facilitated or improved existing human functions, with AIs often applied as specialized tools designed for automation of particular tasks (Dell'Acqua, Kogut, & Perkowski, 2023a; Jia et al., 2023), rather than as active team participants with broad capabilities to generate novel ideas and engage in communication across diverse topics. This distinction highlights the need for a new understanding of how integrating generative AI, with its advanced communicative abilities, influences team interactions, dynamics and outcomes when AI is treated as an equal (Haase & Hanel, 2023), broadly capable partner rather than a narrow tool (Dell'Acqua et al., 2023b; Noy & Zhang, 2023).

Importantly, this oversight limits our understanding of how different approaches to integrating generative AI, capable of contributing diverse ideas and engaging in collaborative efforts, impact team performance and processes. Studies suggest that team performance is not merely the sum of individual abilities but is also shaped by intricate team-level collective factors and dynamics (DeChurch & Mesmer-Magnus, 2010; Mathieu, Maynard, Rapp, & Gilson, 2008; Woolley et al., 2010). The complexity of integrating generative AI's advanced capabilities with the nuances of teamwork, such as fostering effective collaboration, coordination and shared understanding among members (Bittner, Oeste-Reiß, & Leimeister, 2019; Seeber et al., 2020), makes AI integration in teams an intricate yet intriguing challenge to navigate.

To address this gap, our study examines the impact of generative AI on team performance and processes through a preregistered[1], randomized controlled experiment. Participants were assigned to one of three team structures: human-only teams, teams with a single AI, and teams where each member has access to their own AI (multiple-AI condition), to perform two complex professional tasks. This setup allows us to explore whether generative AI enhances team functionality and outcomes, and how different integration structures affect team dynamics (Hemmer et al., 2023; Seeber et al., 2020). A central question is whether incorporating multiple generative AI units into a team leads to an additive increase in performance (Bittner et al., 2019; Choudhary, Marchetti, Shrestha, & Puranam, 2023; Dellermann, Ebel, Söllner, & Leimeister, 2019), or if the dynamics of how the entire team collectively interacts with and utilizes the AI capabilities (Raisch & Krakowski, 2021; Seidel, Berente, Lindberg, Lyytinen, & Nickerson, 2018; Shrestha, Ben-Menahem, & Von Krogh, 2019) are more crucial in determining team-level outcomes. By examining these possibilities, we aim to understand how generative AI's effects scale within teams, providing insights into optimal ways to integrate AI in teams.

---

[1] Available at: https://osf.io/5su8c/?view_only=24f9efec395c4581b880e7773792d23d

While we posit that integrating generative AI in teams enhances overall performance, different AI integration structures may yield nuanced outcomes. On one hand, we postulate that having a single AI unit could foster more focused engagement between the entire team and that AI touchpoint, allowing for collective leveraging of the AI's capabilities. In contrast, providing each team member with their own AI access could distribute the engagement across multiple human-AI dyads, potentially enhancing the diversity of ideas and individual contributions that each member brings to the team. This distinction is crucial for understanding the scalability of generative AI benefits: whether these benefits are additive, with each AI unit contributing incrementally to performance, or dependent on the depth of collective engagement with AI within the team (Choudhary et al., 2023; Raisch & Krakowski, 2021). To explore these questions, we compare the performance of AI-integrated teams against traditional human-only teams and examine how human-AI interactions under different generative AI integration structures influence team outcomes.

To further understand the role of generative AI in team settings and explore whether AI integration could potentially reduce the need for traditional teams, we conducted a second study involving individual-AI pairs. In this study, individuals paired with AI (i.e., a single human participant with a single generative AI touchpoint) were tasked with performing the same tasks as the teams did, using the same evaluation standards. Traditional teams are built to leverage diverse ideas and collaborative efforts (Kozlowski & Ilgen, 2006; Mathieu et al., 2008; Woolley et al., 2010), raising the question of whether generative AI can replicate or even enhance these benefits when used by a single individual. By comparing the performance of individual-AI pairs with both human-only teams and AI-assisted teams, we aim to determine if generative AI can scale up productivity to a level comparable to or exceeding that of team collaboration. This examination provides deeper insights into whether the advantages of AI are primarily rooted in its ability to augment individual capabilities (Grønsund & Aanestad, 2020; Jia et al., 2023; Raisch & Krakowski, 2021) or if the collaborative component of teams (Puranam, 2021; Seeber et al., 2020) remains essential for maximizing outcomes.

Our findings reveal nuanced insights into the role of generative AI in teams. Consistent with prior findings that generative AI improves productivity (Noy & Zhang, 2023), AI-assisted teams consistently outperformed human-only teams across all performance dimensions. However, in our study, we found that including AI in teams explained only 2-4% additional variance in performance outcomes, a much smaller effect size compared to prior studies at the individual level. For instance, Li et al. (2024) and Dell'Acqua et al. (2023) showed that working with AI can explain over 20% to 60% variance in individual productivity. Additionally, working with AI in teams did not significantly reduce task completion time as suggested in individual-level studies (Dell'Acqua et al. 2023; Noy & Zhang, 2023). Yet, linking to research on how AI influences human task satisfaction (Hemmer et al., 2023; Sadeghian & Hassenzahl, 2022), we find that AI integration significantly improved team morale by enhancing members' shared perceptions of confidence and satisfaction compared to traditional human-only teams.

In particular, the structure of AI integration—whether a single AI or multiple AI units—did not result in significant differences in average performance. Teams with multiple AI units did not necessarily outperform those with a single AI, suggesting diminishing returns for increased

AI presence. This result challenges the additive hypothesis, which would predict superior outcomes for teams with more generative AI touchpoints due to the aggregation of individual AI-driven gains. Analyzing human-AI interaction mechanisms, particularly in the multiple-AI condition, revealed that more centralized AI usage—where one or a few team members engaged more deeply with AI—yielded better team outcomes. This finding offers initial evidence that the depth rather than the breadth of AI engagement is more critical for AI-assisted teams.

However, the performance distribution differed significantly between single-AI and multiple-AI teams. Multiple-AI teams exhibited more extreme outcomes, meaning they were more likely to perform either exceptionally well or poorly compared to single-AI teams. This variability suggests that while some teams effectively leverage the additional AI resources (Lebovitz, Lifshitz-Assaf, & Levina, 2022), others struggle with increased coordination demands. Furthermore, teams with higher internal resources, such as high IQ, strong team familiarity, and larger size, were more likely to benefit from multiple AI units, effectively utilizing the potential advantages these additional resources provide.

While our data shows that AI-assisted teams and human only teams outperformed individuals working with generative AI, we also find that individuals working with AI completed tasks significantly faster than teams. Further analysis revealed that one possible reason teams outperformed individuals is that they spent more time on tasks. When individual-AI pairs dedicated similar amounts of time to tasks as teams, their performance matched that of human-only teams but still lagged behind AI-assisted teams. This suggests that while individual-AI pairs can reach the performance levels of traditional teams if they spend sufficient time on tasks, AI-assisted teams still hold an advantage. This underscores the inherent benefits of collaboration that cannot be fully replicated by individuals (Woolley et al., 2010), even when they engage deeply with generative AI (Lebovitz et al., 2022).

In summary, our study seeks to elucidate the impact of generative AI on team performance, exploring whether different AI integration structures enhance outcomes and how these benefits scale within teams. By comparing human-only teams, AI-assisted teams, and individual-AI pairs, we aim to uncover the dynamics of human-AI interactions and their implications for organizational structures. Our findings indicate that while AI-assisted teams outperform human-only teams, the benefits of AI integration are complex and multifaceted. Individual-AI pairs completed tasks in a shorter of amount of time and could match human-only team performance when spending a similar amount of time on the tasks, suggesting potential for reducing reliance on traditional teams. However, considering that AI-assisted teams still outperformed individual-AI pairs, the inherent advantages of collaboration remain significant, highlighting the importance of balancing AI integration with human teamwork to optimize productivity and innovation.

## 2. Research Design and Method

### 2.1 Experiment I

In our first experiment,[2] conducted within a controlled behavioral laboratory setting (see Supplementary Materials[3]), we sought to elucidate the dynamics of generative AI integration and team performance by having participants engage in a series of professional tasks. The study drew on a diverse cohort of 435 working professionals spanning various industries such as marketing, operations, technology research, and general management, in addition to college students. These individuals were randomly allocated to 122 teams, ranging from three to five members, and were tasked with collaboratively completing assignments under varying conditions of AI assistance.

---

[2] The two experiments are approved by IRB (project title: Human Interactions with Artificial Intelligence in Organizations, No. THU-07-2023-02).

[3] Available at: https://osf.io/zq48w/?view_only=ca644891d4234935b1464fbc20ce6e85

The teams were randomly assigned into three conditions: the human-only team condition, where teams operated without AI assistance[4]; the multiple-AI team condition, where the team was provided with as many ChatGPT4.0 touchpoints as there were team members; and the single-AI team condition, where the team was provided with only a single ChatGPT4.0 touchpoint (refer to Figure 1).

[ Insert Figure 1 about here ]

Through the randomized control, 142 participants were assigned to the human-only team, 150 to the single-AI team, and 143 to the multiple-AI team conditions. The number of participants in each of the three conditions for every task remained relatively balanced. For their participation, each individual received compensation valued at $25, offered in a mix of gifts and cash. An incentive was implemented to further encourage their full engagement and effort. We informed participants that the top 20% of teams, based on the quality and creativity of their task outputs, would receive a gift valued at approximately $50.

The experiment encompassed two tasks: a content generating tasks that required participants to create a series of in-depth articles on college students' job hunting and career development, and a strategy development task that had participants develop of a digital transformation strategy for a traditional retail chain. These tasks were designed to entail content creation and innovation as well as strategic thinking. Each task was selected for its relevance, as well as its varying levels of complexity and creativity requirements, thereby providing a comprehensive view of the participants' abilities in collaborating with AI (further task details provided in Supplementary Materials).

---

[4] Incentive is provided to avoid potential negative sentiments among participants in the human-only condition due to not having the opportunity to use AI. Specifically, after human-only teams completed tasks without AI assistance and submitted the tasks, we provided one ChatGPT4.0 touchpoint, allowing the team to further refine their task outcomes (effects of this further refinement not included in our analysis).

Prior to the start of the experiment, participants were required to complete a personal background questionnaire, which collected a wide range of personal data including demographic information, IQ, and previous experience with generative AI. The rationale behind gathering this diverse array of data was to control for the effects of heterogeneity associated with human factors in team-AI collaboration. After completing the initial tasks, participants filled out a post-experiment questionnaire to capture their perceptions towards team processes, including team coordination, potency, information elaboration and satisfaction (see Supplementary Materials). Furthermore, chat log files were extracted from ChatGPT4.0, capturing their interactions with the generative AI, which detailed all prompts and step-by-step responses of participants, providing data for further analysis of the mechanisms underlying human-AI interaction.

To evaluate the performance outcomes of the tasks, we recruited an online panel of judges. Each task output was evaluated based on three dimensions: overall quality, novelty, and usefulness (Amabile, 1983), rated on a scale from one to ten (see Supplementary Materials). The human judges remained blind to both the hypotheses and the conditions under which each task output was created. Each task output was assessed by six independent raters. This approach allowed us to test measurement reliability and to ascertain the level of agreement between raters, resulting in Cronbach's alpha of 0.759, 0.702, and 0.684 across these three dimensions.

In our analysis, we empirically investigated the impact of varying AI implementation conditions on team collaboration using OLS regression analysis. The dataset comprised 244 observational samples from 122 teams, each completing two tasks. We controlled for team characteristics (team age, team gender, team education, team IQ, and team size), team AI usage, and team familiarity. We incorporated task dummies to account for task-specific fixed effects. Additionally, to unify the scales of variables, all control variables were standardized during the regression.

## 2.2 Experiment II

Building on the first experiment, the second experiment sought to assess whether integrating AI could diminish the need for traditional teams. In this phase, individuals paired with AI undertook the same tasks as those performed by teams, adhering to identical evaluation standards. We enlisted a diverse cohort of 139 participants to our behavior lab. Participants completed the same identical questionnaire to gather personal data before the tasks. Each participant then individually worked with ChatGPT4.0 to complete the two tasks.

Each task was appraised by six raters from an online panel similar to Experiment I. The reliability of the evaluations was confirmed with Cronbach's alpha values of 0.770, 0.766, and 0.759 for overall quality, novelty, and usefulness, respectively.

For the analysis, we employed OLS regression on a dataset that included 139 individuals from Experiment II and 122 teams from Experiment I, covering the two tasks. Consequently, the sample size for the empirical regression analysis totaled to 522 observations. In our analysis, we controlled demographic variables (i.e., age, gender, education), IQ, prior experience with generative AI for individual- or team-level, and also the task-specific fixed effect, mirroring the approach in Experiment I.

## 3. Empirical Results

### 3.1 Descriptive Statistics

There was no significant difference in team demographic characteristics across conditions, with the exception of average age, confirming the effectiveness of our randomization process (refer to Table 1 and Table 2 in Appendix A). However, the analysis of task outcomes across three experimental conditions reveals notable differences in performance. Teams equipped with AI assistance, either as a single AI or multiple AIs, consistently outperformed teams without AI assistance. Specifically, single-AI team and multiple-AI team conditions demonstrated superior performance in overall quality, with mean scores of 5.994 and 5.968, respectively, compared to 5.591 for the human-only team condition. Similarly, in terms of novelty and usefulness, AI-assisted teams exhibited higher mean scores, indicating that the integration of AI into teams enhances their ability to generate innovative and practical solutions. As there was no significant difference between single-AI team and multiple-AI team, the difference in their impact on task performance was not as pronounced as expected. Figure 2 provides a more intuitive visual representation of these results.

[ Insert Figure 2 about here ]

**3.2 Empirical Results**

**3.2.1 Generative AI Integration and Team Performance**

Figure 3 presents the regression coefficients and 95% confidence intervals for AI integration on team performance, including the human-only condition for comparison. The results demonstrate a pronounced effect of AI integration on various dimensions of task performance. Specifically, as shown in Table 1 in Appendix B, Model 2 indicates a significant enhancement in overall quality when AI is utilized ($\beta = 0.417$, $p < .01$), explaining additional 3.7% variance in performance. Models 3 and 5 reveal that AI positively influences both the novelty ($\beta = 0.344$, $p < .05$) and usefulness ($\beta = 0.250$, $p < .10$), explaining an additional 2.5% and 1.4% variance in outcomes, respectively. This aligns with the visual evidence provided in Figure 2, which illustrates the regression coefficients of the AI integration effect.

[ Insert Figure 3 ]

### 3.2.2 Different AI Integration Structures and Team Performance

Figure 4 presents the regression coefficients and 95% confidence intervals for AI integration structures—particularly the use of single versus multiple AIs—on team performance. Contrary to initial expectations, the empirical results demonstrate that the introduction of multiple AIs into team settings does not significantly enhance task performance. Specifically, as shown in Table 3 in Appendix B, our analysis reveals that the presence of multiple AIs did not substantially change the overall quality of outcomes ($\beta = -0.020$, $p > .10$), and while there was a slight increase in novelty ($\beta = 0.052$, $p > .10$) and a more noticeable improvement in usefulness ($\beta = 0.092$, $p > .10$), these effects did not achieve statistical significance at conventional levels. This suggests that the benefits of AI assistance do not scale additively with the number of generative AI units integrated into a team (refer to Table 2 in Appendix B).

We quantified the observed null effect using a Bayesian approach. We employed Bayesian linear regression with default prior distribution. The results in Table 4 in Appendix B showed that there is no significant difference in task performance between teams with a single AI and those with multiple AIs (*Mean* = -0.072, CI = [-0.303, 0.169] for *Overall*; *Mean* = 0.025, CI = [-0.220, 0.262] for *Novelty*; *Mean* = 0.076, CI = [-0.184, 0.323] for *Usefulness*). We also established a baseline model that did not include the variable "multiple-AI" for comparative analysis. For the Overall dimension, the inverse Bayes Factor of the difference between the two models was 1355. This indicates that in our sample, it is 1355 times more likely that the "multiple-AI" variable did not contribute significantly to improvements in the overall dimension. Similarly, for the novelty and usefulness dimensions, the inverse Bayes Factors of the differences between the models were 1277 and 1261, respectively. Collectively, these results provide strong evidence supporting the null effect of multiple-AI.

[ Insert Figure 4 about here ]

This lack of significant performance differences between single and multiple-AI team conditions indicates that generative AI usage may not having additive scaling in teams. While simply increasing the number of generative AI touchpoints does not automatically translate to better outcomes, there may be more nuanced factors, such as the depth of engagement with the AI, that determine whether and how teams can capitalize on having multiple AI units at their disposal.

### 3.2.3 Relative Distribution of Single-AI Team and Multiple-AI Team

To examine the lack of significant differences in average performance between single-AI teams and multiple-AI teams, as well as explore the potential existence of interaction depth-based mechanisms, we delved deeper into the relative performance distributions of these two conditions. Figure 5 displays the relative probability density functions for the task performance of single-AI teams versus multiple-AI teams. When the relative density value exceeds 1, it indicates a higher frequency of occurrences at that quantile for multiple-AI teams; conversely, values less than 1 indicate a lower frequency.

The analysis revealed notable distinctions in the performance distributions between single-AI and multiple-AI teams. Specifically, multiple-AI teams demonstrated more extreme outcomes, suggesting a greater likelihood of either exceptionally high or exceptionally low performance compared to single-AI teams. This variability is highlighted by the 95% confidence intervals marked on the relative probability density function across various quantiles in the graph.

This pattern suggests that while having multiple AI touchpoints may not consistently enhance performance metrics, it does introduce greater variability in outcomes. This finding underscores the complexity and potential risks associated with managing multiple AI systems within team settings. As such, understanding what factors enable teams to effectively leverage multiple AI capabilities becomes crucial for unlocking the potential benefits while managing the added complexity.

[ Insert Figure 5 about here ]

### 3.2.4 Heterogeneity for AI teams

In our examination of heterogeneity effects within AI-assisted teams, we aimed to determine if certain team characteristics could amplify the efficacy of multiple AI units. Figure 6 showcases margin plots with 95% confidence intervals, illustrating the performance outcomes from the interaction between internal resource variables—such as team IQ, team familiarity, and team size—and multiple AI integrations. The findings indicate that teams endowed with higher human-based resources—namely higher IQ levels, stronger intra-team familiarity, and larger sizes—are more adept at benefiting from the incorporation of multiple AI touchpoints.

Specifically, as detailed in Table 5 in Appendix B, the interaction between multiple AI units and team IQ had a notable positive impact on *Overall* ($\beta = 0.279$, $p < .10$) and *Usefulness* ($\beta = 0.243$, $p < .10$), suggesting that teams with superior cognitive resources are better equipped to harness the capabilities of multiple AIs. Likewise, the interaction with team size was significant, enhancing *Overall* ($\beta = 0.288$, $p < .05$), *Novelty* ($\beta = 0.235$, $p < .10$), and *Usefulness* ($\beta = 0.252$, $p < .10$), indicating that larger teams could more effectively utilize additional AI touchpoints to boost these performance metrics. Furthermore, the interaction with team familiarity also yielded positive effects on *Overall* ($\beta = 0.248$, $p < .10$), *Novelty* ($\beta = 0.275$, $p < .10$), and *Usefulness* ($\beta = 0.278$, $p < .05$), underscoring the advantage that well-acquainted teams have in integrating and leveraging multiple AI units.

These insights emphasize the critical role of team composition and internal dynamics in the successful integration of multiple generative AI touchpoints into team settings. Teams with robust internal capacities are shown to exploit the potential benefits of multiple AI units more effectively, suggesting that integration strategies should be customized to align with the specific resources and characteristics of the team.

[ Insert Figure 6 about here ]

### 3.2.5 Team Performance Versus Individual-AI Pair Performance

In the individual-AI pair condition, the task performance of *Overall*, *Novelty*, and *Usefulness* were recorded as 5.219, 5.056, and 5.228, respectively (refer to Table 3 in Appendix A). As demonstrated in Figure 7, their performance distribution significantly lags behind that of both AI teams (single & multiple AI teams) and human-only teams.

Figure 8 illustrates the regression coefficients and 95% confidence intervals comparing the performance of individual-AI pairs with AI teams (single & multiple) and human-only teams, setting the human-only team condition as the baseline for comparison in Experiment II. The results indicate significant disparities in performance among these conditions. Specifically, as detailed in Table 6 in Appendix B, individual-AI pairs showed a notable underperformance across all metrics, with negative impacts on *Overall* ($\beta = -0.378$, $p < .05$), *Novelty* ($\beta = -0.410$, $p < .01$), and *Usefulness* ($\beta = -0.438$, $p < .01$). This suggests that individual-AI pairs are less effective than both human-only and AI-enhanced team settings. The effects between AI teams (single & multiple) are similar to those in Figure 2.

[ Insert Figure 7 & Figure 8 about here ]

Further, we also investigated the relative performance distributions among individual-AI pair, human-only team, and AI team conditions. Figure 1 in Appendix B revealed that individual-AI pairs demonstrated more extreme outcomes than human-only teams, indicating a higher variance in performance. Figure 2 in Appendix B further illustrated that individual-AI pairs exhibited a significantly greater likelihood of exceptionally low performance when compared to AI teams, for both single- and multiple-AI configurations. Overall, these findings suggest that individuals paired with AI, despite the potential for high achievement, are more likely to experience low performance frequencies compared to teams. This emphasizes the critical role of collaborative dynamics in harnessing the full potential of generative AI.

### 3.2.6 Task Completion Time for Different Conditions

In addition to task performance, we also examined task efficiency across different conditions, with task completion time serving as the dependent variable. As detailed in Table 7 in Appendix B, we first compare the differences in task completion time between AI teams and human-only teams. The results indicate no significant difference in task completion time between AI teams and human-only teams ($\beta = 0.037$, $p > .10$). This observed null effect was re-examined using Bayesian linear regression, which provided strong evidence supporting the null effect of AI teams (inverse Bayes factor = 1333) (for details, refer to Table 8 in Appendix B).

Next, we compared the completion time differences between multiple-AI teams and single-AI teams. The results suggest that multiple-AI teams take more time to complete tasks, indicating relatively lower efficiency compared to single-AI teams (Model 2, $\beta = 0.128$, $p < .10$).

Finally, we examined the individual-AI pair condition, using the human-only team condition as the baseline for comparison. Compared to the human-only team condition, the individual-AI pair condition demonstrated significantly higher efficiency, completing tasks in less time (Model 3, $\beta$ = -0.659, $p$ < .01). This suggests that while individual-AI pairs may not consistently match the collaborative performance levels of teams, they can offer advantages in task completion speed.

Building on the above findings, we hypothesized that one possible reason teams outperformed individuals might be the additional time teams spent on tasks. Consequently, we posed a new question: if individual-AI pairs were allotted the same amount of time on tasks as teams, could their performance match or even surpass that of the teams? Specifically, we control task completion time based on the model above in Table 6. As shown in Table 9 in Appendix B, individual-AI pairs show no significant impacts on *Overall* ($\beta$ = -0.153, $p$ > .10), *Novelty* ($\beta$ = -0.173, $p$ > .10), and *Usefulness* ($\beta$ = -0.188, $p$ > .10). Similarly, this observed null effect was re-examined using Bayesian linear regression, which provide strong evidence supporting the null effect of individual-AI pairs (inverse Bayes factor is 263 for *Overall*, 714 for *Novelty*, 909 for *Usefulness*) (for details, refer to Table 10 in Appendix B). While AI teams (single- and multiple-AI conditions) still shows significant positive impacts on *Overall* ($\beta$ = 0.422, $p$ < .01), *Novelty* ($\beta$ = 0.372, $p$ > .05), and *Usefulness* ($\beta$ = 0.251, $p$ < .10).

Figure 8 visually illustrates the comparison results between team performance and individual-AI pair performance, both with and without controlling for task completion time. These findings suggest that when time investment is held constant, individual-AI pairs are able to achieve performance levels comparable to those of traditional teams. However, they do not reach the performance levels of teams that are enhanced with generative AI. This underscores the importance of collaborative dynamics in fully leveraging the effectiveness of AI.

Employing Coarsened Exact Matching (CEM) with task completion time as a covariate and adhering to the K2K matching principle, we also created matched samples where each individual-AI pair was paired with a human-only team or an AI team (single- and multiple-AI conditions) that completed the specific task in the same amount of time. This approach facilitated the construction of a new empirical dataset. The results above are similar robust (details shown in Table 11 and Table 12 in Appendix B).

### 3.2.7 Mechanisms and Team-AI Collaboration Tactics

**(1) Interaction Indicators.** To gain a deeper understanding of team-AI collaboration, we analyzed the interaction logs that capture every prompt and response between team members and the generative AI. We measured the team members' total prompt length and the total interaction rounds. Human input was also assessed utilizing the GPT-4.0 API, known for its robust text understanding capabilities (Dell'Acqua et al., 2023). From a depth-based perspective, the most impactful human-AI interactions may stem from the deepest engagement, where team members leverage AI's capabilities to the fullest extent. Therefore, to capture the effect of deep engagement, we selected only the top 25% of prompts by score to construct a human input indicator (detailed methods and measurements can be found in Supplementary Materials).

Regression results indicated that the total prompt length, total interaction rounds, and GPT-assessed human input all significantly and positively predicted task performance (see Table 13 in Appendix B, where interaction indicators were all standardized). Additionally, we conducted robustness checks by considering the average interaction rounds and average prompt length of team members. For human input, we also constructed indicators using the top 50% of prompts (see Table 14 in Appendix B). All of the results remained robust.

**(2) Interaction Patterns.** We further explored the collaboration patterns among different AI touchpoints within multiple-AI teams. Based on the prompt length and interaction rounds of each team member, we calculated the Gini coefficient to reflect the division of contributions within the team. A higher Gini coefficient indicates a more uneven distribution of contribution levels within the team, signifying that one or a few team members engaged more deeply with AI. As shown in Table 15 in Appendix B, the Gini coefficient for prompt length significantly positively predicted team performance, whereas the Gini coefficient for interaction rounds did not show a significant impact. Overall, the results revealed that more centralized AI usage—where one or a few team members engaged more deeply with AI—yielded better team outcomes.

### 3.2.8 Team Process Perception

Finally, we examined team process perceptions across different conditions. The results, as detailed in Table 16 in Appendix B, indicate that compared to human-only teams, AI teams (single- and multiple-AI conditions) had varying impacts on these team processes. Specifically, AI teams significantly enhanced team potency ($\beta = 0.313$, $p < .05$) and team satisfaction ($\beta = 0.266$, $p < .10$), while their effects on team coordination ($\beta = 0.010$, $p > .10$) and team information elaboration ($\beta = 0.105$, $p > .10$) were not significant. We further examined the differences between single and multiple AI teams. According to the results in Table 17 in Appendix B, no significant differences were found in team dynamics.

These findings underscore that generative AI integration significantly boosts team morale by enhancing members' shared perceptions of confidence and satisfaction compared to traditional human-only teams. However, the additional integration of multiple AIs, as compared to a single AI, does not yield significant differences in the evaluated team processes. This suggests that the benefits of generative AI in team dynamics may not additively increase with the number of AIs involved, highlighting the effectiveness of AI-enhanced teamwork without necessarily scaling up the number of AI elements.

**4. Discussion**

The findings from these experiments offer compelling insights into the dynamics of human-AI collaboration and its impact on team performance. The results demonstrate that integrating generative AI into teams can significantly enhance their output quality, novelty, and usefulness compared to traditional human-only teams. This effect holds true regardless of whether teams have access to a single AI or multiple AI touchpoints, suggesting that the mere presence of generative AI augmentation is a key driver of performance gains. Notably, the effect size for generative AI's impact on teams is much smaller than that for individual productivity reported in previous studies, suggesting the complexity of integrating AI in teams (Dell'Acqua et al., 2023; Li et al., 2024; Noy & Zhang, 2023).

Additionally, the study also revealed that the addition of multiple AI touchpoints within a team did not yield significant differences in team performance compared to teams with a single AI touchpoint. This suggests that the benefits of generative AI integration on team processes may plateau after a certain point, and that scaling up the number of AI units within a team may not necessarily lead to further improvements. Further analysis of the interaction patterns within teams revealed that more centralized AI usage, where one or a few team members engaged deeply with the AI, yielded better team outcomes. This finding suggests that effective AI integration may not necessarily require every team member to extensively interact with the AI, but rather is supported by having dedicated "AI specialists" within the team who can leverage the technology's capabilities to the fullest extent.

Furthermore, results from the post-hoc study showed that individual-AI pairs were able to achieve a similar performance level as human-only teams when using the same amount of time to complete tasks. This suggests that having individuals leverage generative AI can potentially reduce the need for traditional teams in certain scenarios. In particular, generative AI can serve as a team member by providing ideas and sharing the workload, thereby diminishing the necessity for multiple human collaborators. However, individual-AI pairs still underperformed compared to AI-enhanced teams, even when controlling for time spent on tasks. This underscores that while generative AI can augment individual capabilities, the collaborative dynamics and diverse perspectives present in teams remain essential for fully optimizing outcomes and harnessing AI's potential.

The synergistic interplay between human team members and generative AI appears crucial for consistently achieving high-quality outputs that surpass what individuals or human-only teams can produce. Interestingly, the integration of generative AI into teams not only impacted task performance but also influenced team dynamics and perceptions. AI-enhanced teams reported significantly higher levels of team potency and satisfaction compared to human-only teams.

Overall, these findings have significant implications for the future of work and the role of AI in organizational settings. The ability to augment human teams with generative AI presents a promising avenue for enhancing productivity, creativity, and strategic decision-making. However, it is crucial to carefully consider the dynamics of human-AI collaboration and to design effective strategies for integrating AI into teams in a way that maximizes its potential while preserving the benefits of human collaboration.

**Figure 1 Experimental Design for Experiment I**

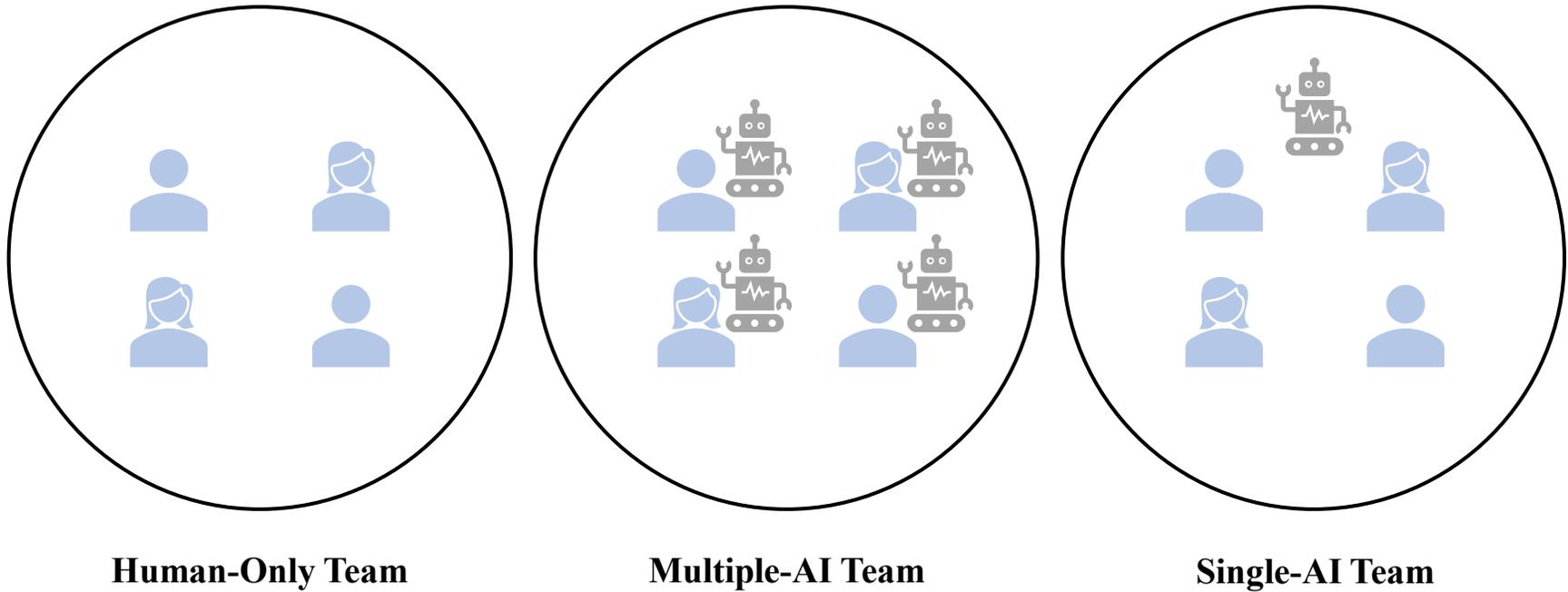

**Human-Only Team**  **Multiple-AI Team**  **Single-AI Team**

Note: Teams were randomly assigned into three conditions: the human-only team condition (142 participants), the multiple-AI team condition (143 participants, with as many ChatGPT 4.0 touchpoints as team members), and the single-AI team condition (150 participants, with only a single ChatGPT 4.0 touchpoint).

**Figure 2 Performance Distribution for Experiment I**

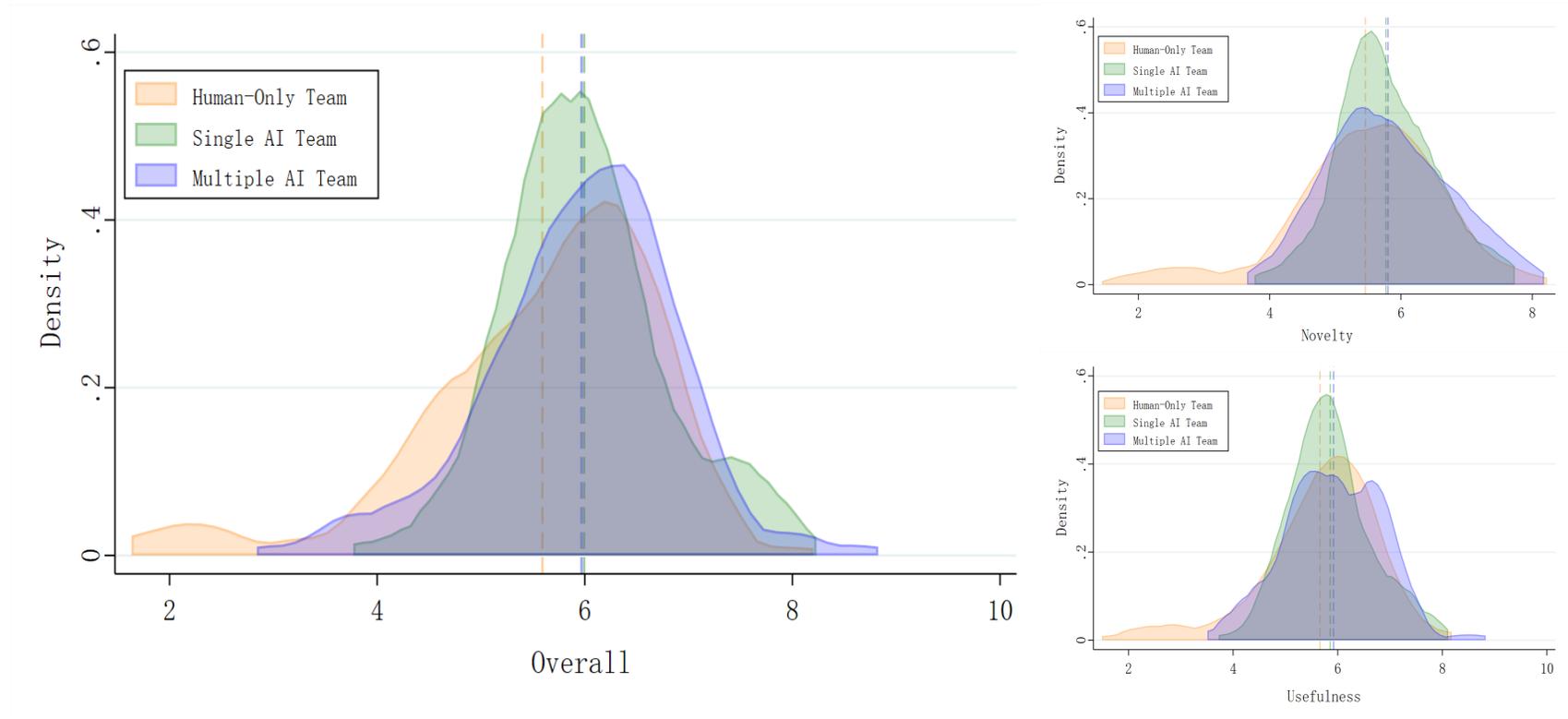

Note: The chart displays the distribution of team performance under different conditions. In the chart, orange represents human-only teams, green represents single-AI teams, and blue represents multiple-AI teams. The vertical lines indicate the average performance scores for each condition. And the left side shows the scores for the Overall dimension, while the right side, from top to bottom, shows the scores for the Novelty and Usefulness dimensions.

**Figure 3 Generative AI Integration and Team Performance (Experiment I, including AI team and human-only team conditions)**

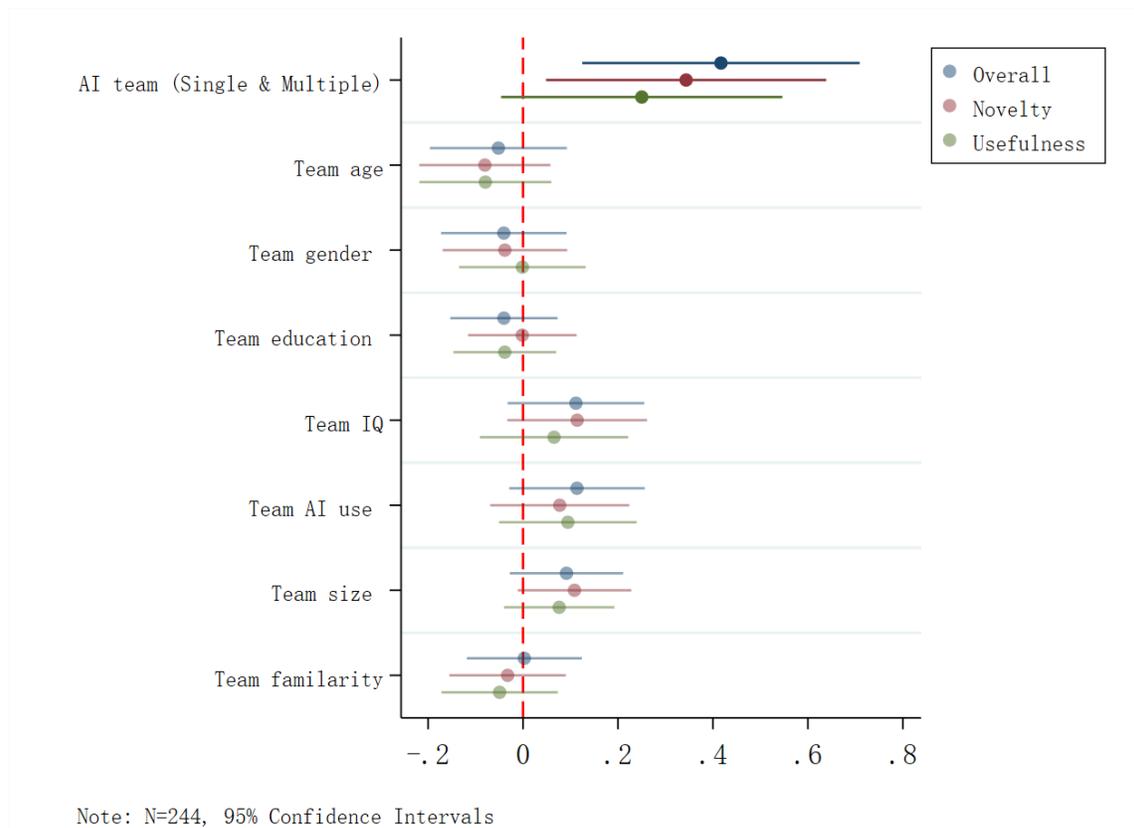

Note: The chart shows the coefficient plot of our regression analysis, using the human-only teams as the reference group. It employs an OLS regression model, controlling for task fixed effects, and all control variables have been standardized. In the chart, blue represents the Overall dimension, red represents the Novelty dimension, and green represents the Usefulness dimension. The dots indicate regression coefficients. The red vertical line at zero value serves as a reference line. The horizontal lines represent 95% confidence intervals.

**Figure 4 Generative AI Integration and Team Performance (Experiment I, only AI teams)**

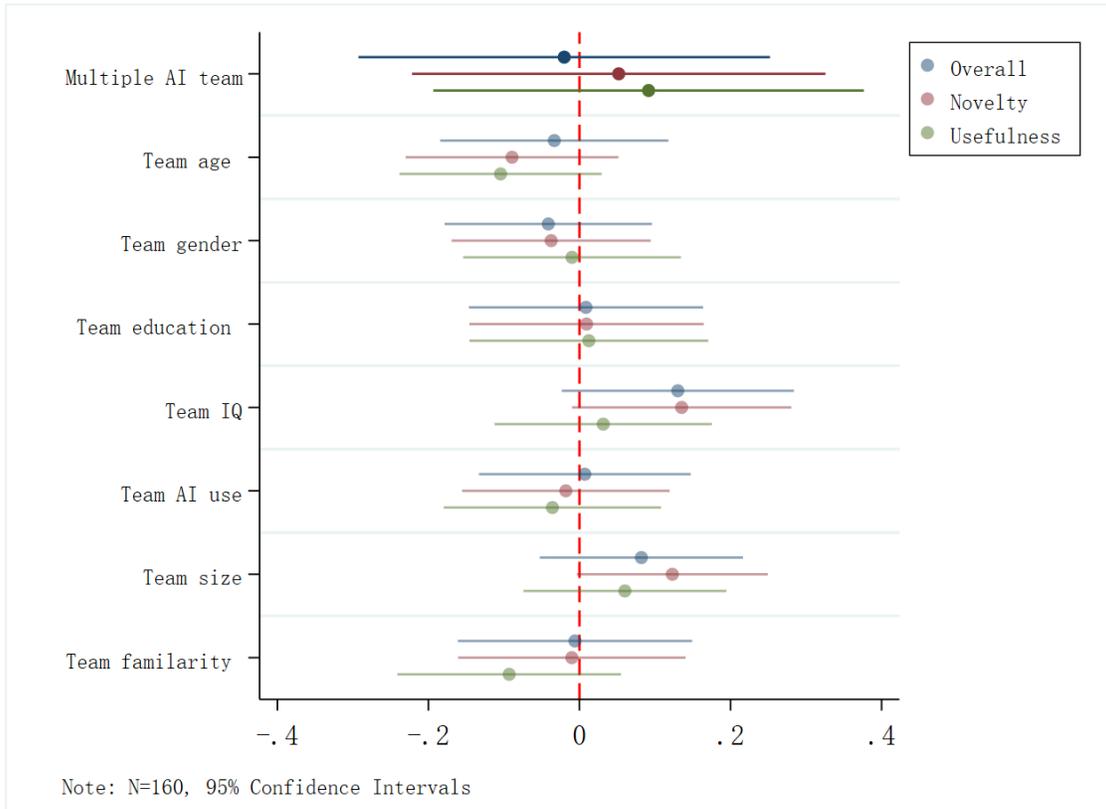

Note: The chart shows the coefficient plot of our regression analysis with only AI-team samples, using the single-AI teams as the reference group. It employs an OLS regression model, controlling for task fixed effects, and all control variables have been standardized. In the chart, blue represents the Overall dimension, red represents the Novelty dimension, and green represents the Usefulness dimension. The dots indicate regression coefficients. The red vertical line at zero value serves as a reference line. The horizontal lines represent 95% confidence intervals.

**Figure 5 Relative Distribution of Single-AI Team and Multiple-AI Team**

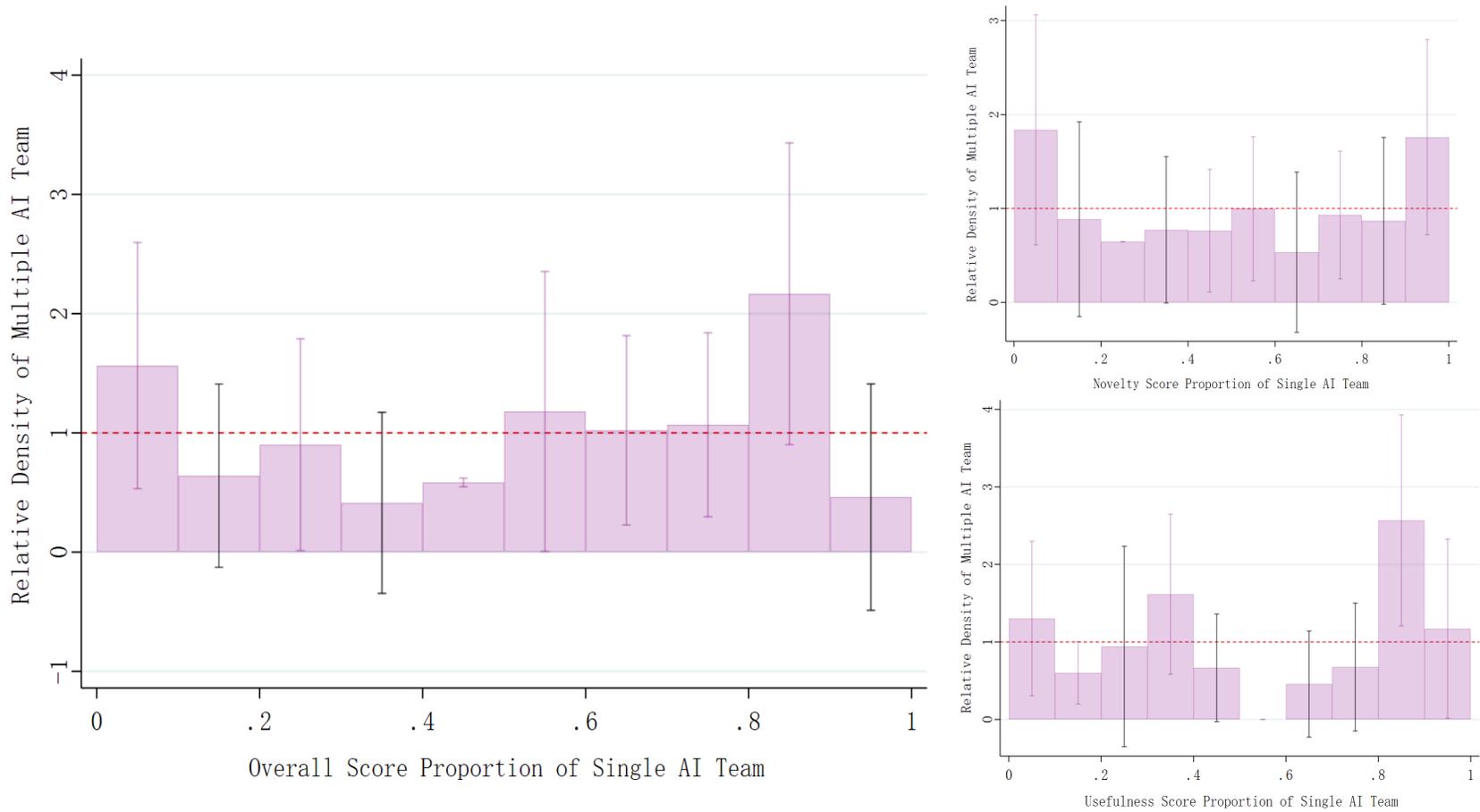

Note: The chart displays the relative distribution of single-AI teams and multiple-AI teams on team performance, with the single-AI teams as reference group. The left side shows the scores for the Overall dimension, while the right side, from top to bottom, displays scores for the Novelty and Usefulness dimensions. The red horizontal line at the value of 1 serves as a reference line. When the relative density value exceeds 1, it indicates a higher frequency of occurrences at that quantile for multiple-AI teams; conversely, values less than 1 indicate a lower frequency. The vertical lines represent the 95% confidence intervals for relative density values at specific score proportion intervals. Grey vertical lines indicate intervals that include 0, not reaching a significant level.

**Figure 6 Heterogeneity effects (Experiment I, only AI team)**

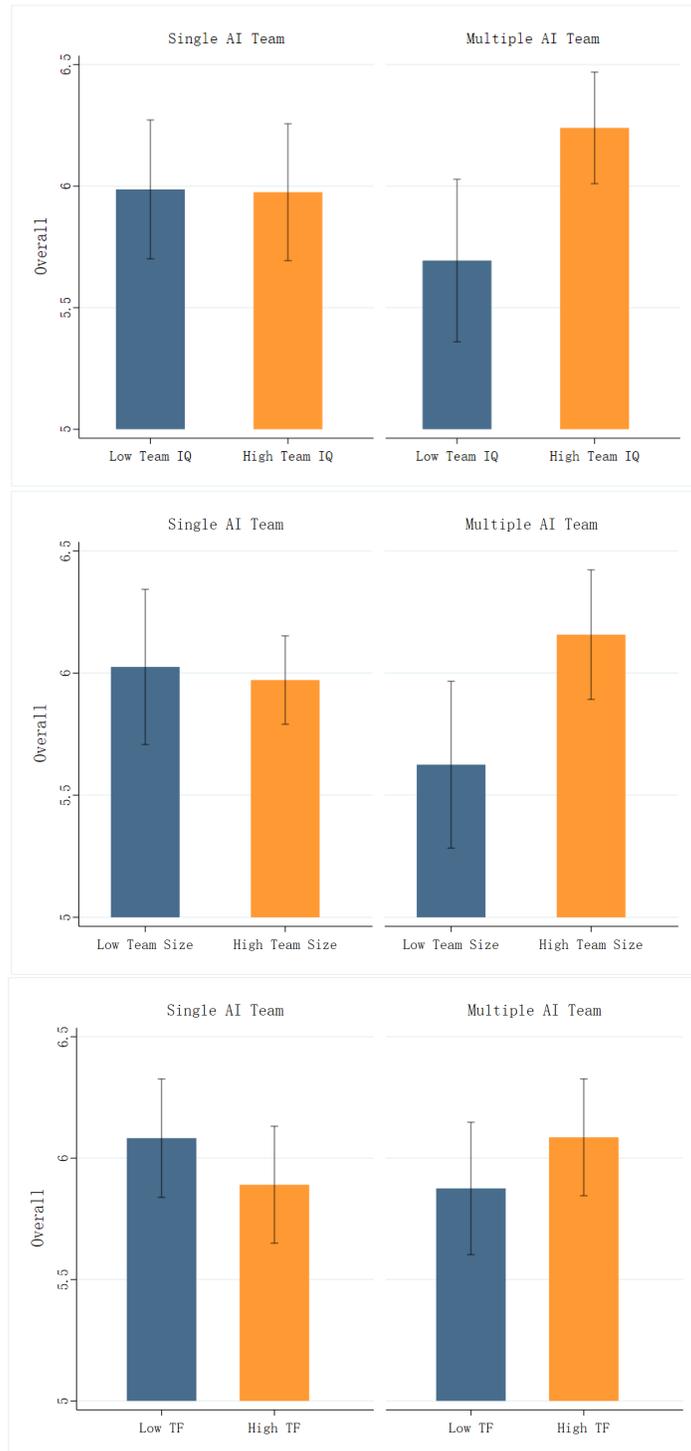

Note: a) TF represent team familiarity, N=160. b) The chart above displays margins plots with team IQ, team size, and team familiarity as moderating variables. The left side of each set of graphs shows the margins plot for the Overall dimension, while the right side, from top to bottom, displays the margins plots for the Novelty and Usefulness dimensions. The bar graphs represent specific scores when the moderating variable takes different values (mean +/- SD) under specific experimental conditions (single-AI team or multiple-AI team). The vertical lines represent their 95% confidence intervals. For team IQ and team familiarity, the interaction is significant at $p < .10$, and for team size, the interaction is statistically significant at $p < .05$.

**Figure 7 Performance Distribution for Experiment II**

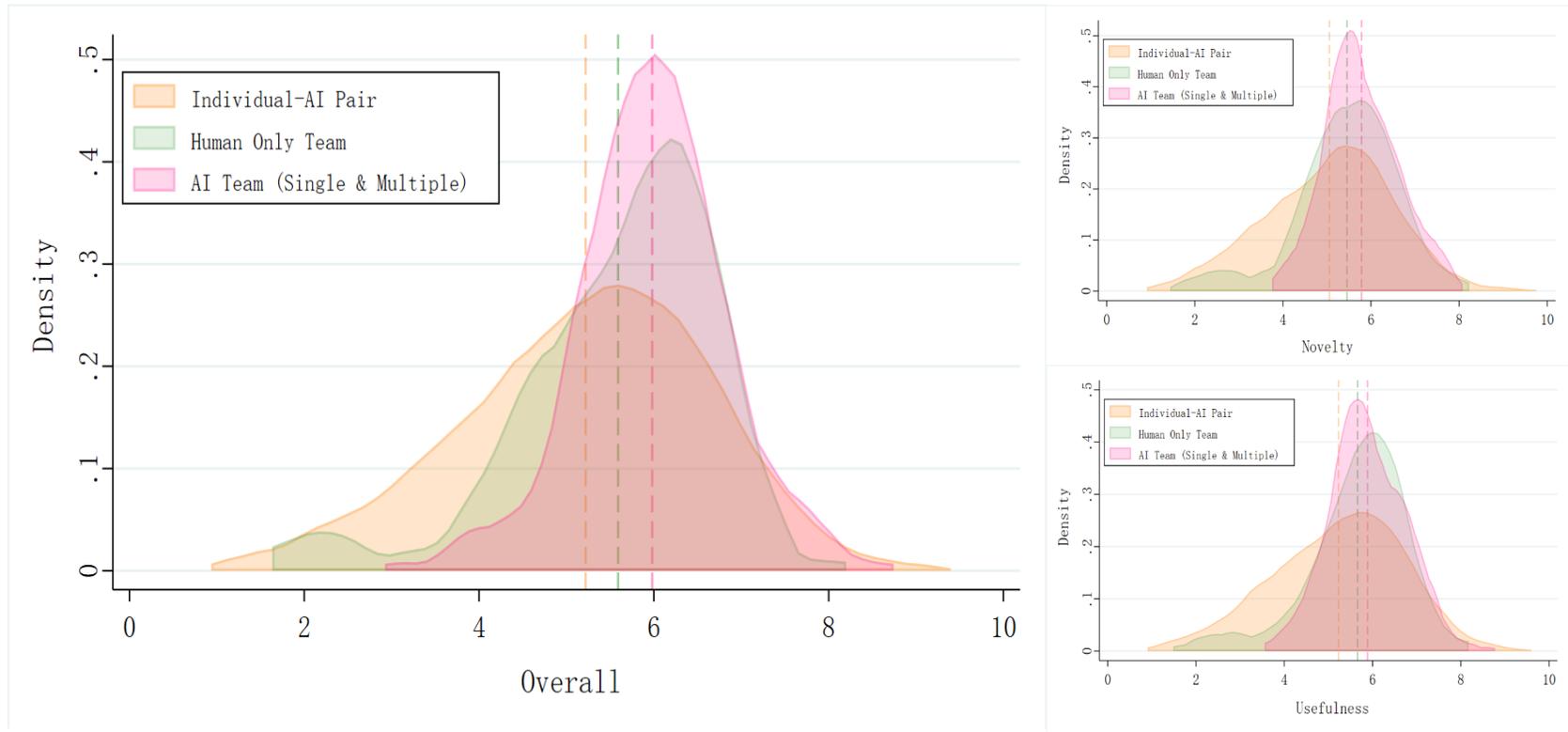

Note: The chart displays the distribution of team performance under different conditions. In the chart, orange represents individual-AI pairs, green represents human-only teams, and pink represents AI teams (single & multiple). The vertical lines indicate the average performance scores for each condition. And the left side shows the scores for the Overall dimension, while the right side, from top to bottom, shows the scores for the Novelty and Usefulness dimensions.

**Figure 8 Team Performance VS Individual-AI Pair Performance (Experiment II)**

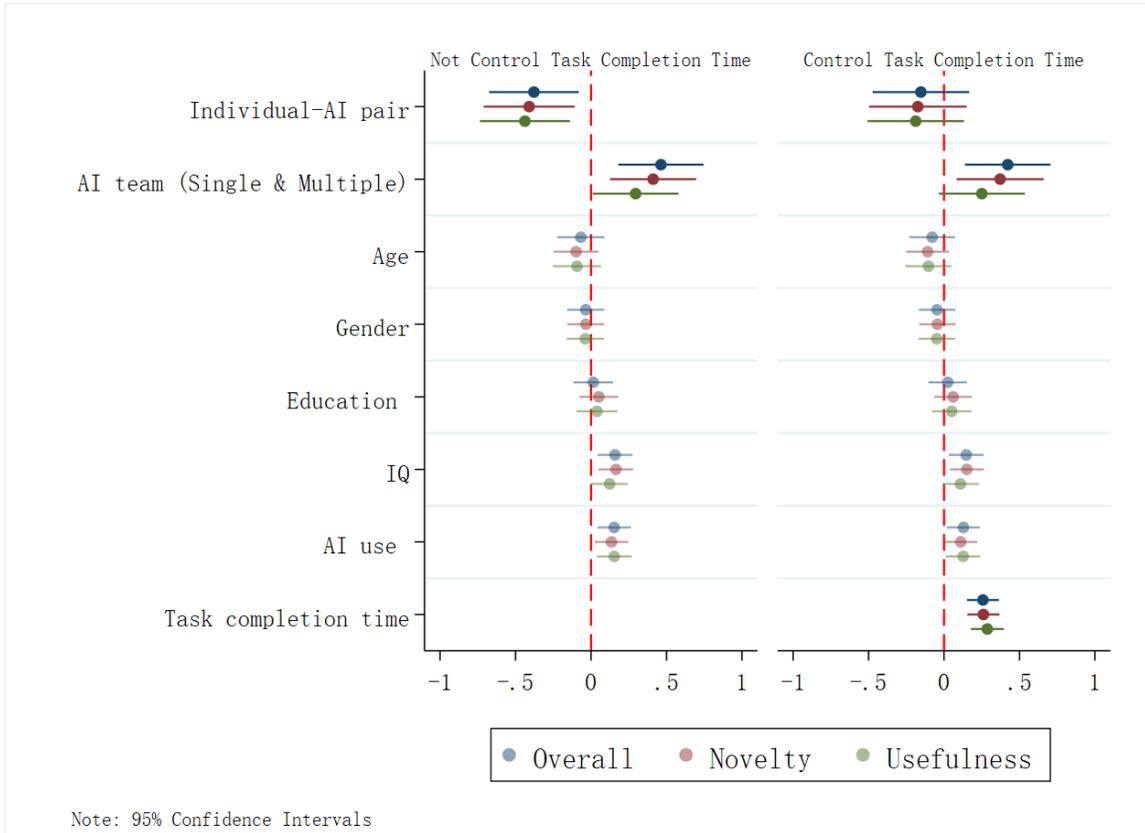

Note: The chart displays the coefficient plot of our regression analyses, using the human-only teams as the reference group. The left side shows the regression results without controlling for task completion time, while the right side shows the results with task completion time controlled. The two model both employ an OLS regression model, controlling for task fixed effects, and all control variables have been standardized. In the chart, blue represents the Overall dimension, red represents the Novelty dimension, and green represents the Usefulness dimension. The dots indicate regression coefficients. The red vertical line at zero value serves as a reference line. The horizontal lines represent 95% confidence intervals.

# Appendix A

**Descriptive Statistics of Experiment I**

Table 1 provides a descriptive summary of the primary variables for Experiment I. The teams have an average age of 25.774, with males constituting 34.3% of the sample. The level of team education is high, with an average of 3.691, indicating that the average education level of team members is above a bachelor's degree. The average AI usage by teams is 0.813, indicating that, on average, 81.3% of team members have experience in using generative AI. Additionally, the value for team familiarity is 0.132, indicating a relatively low proportion of team members knowing each other.

**Table 1 Descriptive Statistics (Experiment I)**

| | Variable | Human-Only Team (N=42) | Single-AI Team (N=41) | Multiple-AI Team (N=39) |
|---|---|---|---|---|
| | | Mean (SD) | Mean (SD) | Mean (SD) |
| Team Characteristics | Team age | 24.646 (3.155) | 25.653 (3.446) | 27.116 (3.764) |
| | Team gender | 0.382 (0.303) | 0.306 (0.248) | 0.339 (0.232) |
| | Team education | 3.633 (0.689) | 3.728 (0.525) | 3.714 (0.389) |
| | Team IQ | 15.808 (1.519) | 15.265 (1.552) | 15.428 (1.549) |
| | Team AI use | 0.826 (0.237) | 0.831 (0.207) | 0.779 (0.214) |
| | Team size | 3.381 (0.582) | 3.659 (0.728) | 3.667 (0.577) |
| | Team familiarity | 0.179 (0.304) | 0.112 (0.200) | 0.103 (0.190) |
| Task Outcomes | | Human-Only Team (N=84) | Single-AI Team (N=82) | Multiple-AI Team (N=78) |
| | | Mean (SD) | Mean (SD) | Mean (SD) |
| Total | Overall | 5.591 (1.139) | 5.994 (0.778) | 5.968 (0.910) |
| | Novelty | 5.456 (1.146) | 5.770 (0.729) | 5.801 (0.913) |
| | Usefulness | 5.661 (1.126) | 5.856 (0.771) | 5.919 (0.910) |
| Task1 | Overall | 5.833 (0.978) | 6.008 (0.574) | 6.030 (0.796) |
| | Novelty | 5.679 (1.160) | 5.724 (0.667) | 5.932 (0.966) |
| | Usefulness | 5.881 (1.095) | 5.850 (0.569) | 5.910 (0.864) |
| Task2 | Overall | 5.349 (1.244) | 5.980 (0.947) | 5.906 (1.017) |
| | Novelty | 5.234 (1.101) | 5.817 (0.792) | 5.671 (0.849) |
| | Usefulness | 5.440 (1.126) | 5.862 (0.939) | 5.927 (0.965) |

*Note: Team age* represents the average age of team members. *Team gender* measures the proportion of male members in the team. *Team education* represents the average education levels of team members. And team members' education levels are rated on a scale from 1 to 5, representing high school or below, associate degree, bachelor's degree, master's degree, and doctoral degree, respectively. *Team IQ* represents the average *IQ* level of the team: The *IQ* levels of team members are derived from 18 *IQ* test items (Sefcek et al., 2016), scaled from 0 to 18. *Team AI use* measures the

average AI use of the team. AI use is represented by a binary variable, where a value of 1 indicates that generative AI has been used at least once, and a value of 0 signifies that generative AI has never been used. *Team size* reflects the total number of members in the team. *Team familiarity* is evaluated by each member rating their familiarity of other members on a scale of 0 (do not know) to 1 (know). Following the measurement approach for valued relations set forth by Sparrowe and colleagues (2001), the familiarity level of the team is calculated by summing all the familiarity ratings among team members and then dividing by the total number of relationships among team members.

In Table 2, ANOVA analysis of team characteristics across the three experimental conditions (i.e., human-only, single-AI, and multiple-AI conditions) revealed no significant differences in these characteristics, except for team age between human-only condition and multiple-AI condition. This confirms the effectiveness of our randomization.

**Table 2 ANOVA Analysis for Experiment I**

| | Variable | ANOVA | | |
|---|---|---|---|---|
| | | Human-Only Team VS Single-AI Team | Single-AI Team VS Multiple-AI Team | Human-Only Team VS Multiple-AI Team |
| Team Characteristics | Team age | 1.007 | -1.463 | 2.469*** |
| | Team gender | -0.076 | -0.033 | -0.043 |
| | Team education | 0.095 | 0.014 | 0.081 |
| | Team IQ | -0.543 | -0.163 | -0.38 |
| | Team AI use | 0.005 | 0.052 | -0.047 |
| | Team size | 0.278 | -0.008 | 0.286 |
| | Team familiarity | -0.0668 | 0.009 | -0.076 |

**Descriptive Statistics of Experiment II**

Table 3 presents a descriptive summary of the primary variables for the individual participants in Experiment II. The average age of the participants is 23.41, with males comprising 29.5% of the sample. The participants exhibit a high level of education, with an average score of 3.468, suggesting that the average educational attainment exceeds a bachelor's degree. Participants' average IQ of 15.91 is slightly higher than the average of participants in Experiment I. The average AI usage score is 0.763, indicating that approximately 76.3% of the individual participants have experience using generative AI. These variables are fundamentally similar to the individual characteristics in Experiment

I. As for the task outcomes, we can see that the results of individual-AI pair are notably lower than the team's performance outputs in Experiment I across all dimensions including overall, novelty and usefulness.

**Table 3 Descriptive Statistics (Experiment II)**

| | Variable | Individual-AI Pair (N=139) | | |
|---|---|---|---|---|
| | | Mean (SD) | Min | Max |
| Team Characteristics | Age | 23.41 (4.541) | 18 | 44 |
| | Gender | 0.295 (0.458) | 0 | 1 |
| | Education | 3.468 (0.764) | 1 | 5 |
| | IQ | 15.91 (1.715) | 12 | 18 |
| | AI use | 0.763 (0.427) | 0 | 1 |
| Task Outcomes | | Individual-AI Pair (N=278) | | |
| | | Mean (SD) | Min | Max |
| Total | Overall | 5.219 (1.421) | 1.333 | 9 |
| | Novelty | 5.056 (1.435) | 1.333 | 9.333 |
| | Usefulness | 5.228 (1.466) | 1.333 | 9.167 |
| Task1 | Overall | 5.194 (1.397) | 1.333 | 8.5 |
| | Novelty | 5.019 (1.429) | 1.333 | 9.333 |
| | Usefulness | 5.195 (1.467) | 1.333 | 9.167 |
| Task2 | Overall | 5.243 (1.449) | 1.667 | 9 |
| | Novelty | 5.092 (1.444) | 1.5 | 8.667 |
| | Usefulness | 5.26 (1.469) | 1.667 | 8.667 |

*Note:* Variables consist with those at team level. Age is recorded in years. Gender is categorized with a binary variable, with males assigned a value of 1. Education is ranged on a scale from 1 to 5, representing high school or below, associate degree, bachelor's degree, master's degree, and doctoral degree, respectively. IQ, derived from 18 IQ test items (Sefcek, et al., 2016), were scaled from 0 to 18. AI use is represented by a binary variable, where a value of 1 indicates that generative AI has been used at least once, and a value of 0 signifies that generative AI has never been used.

# Appendix B

**Table 1 Generative AI Integration and Team Performance (Experiment I, including AI team and human-only team conditions)**

|  | (1) *Overall* | (2) *Overall* | (3) *Novelty* | (4) *Novelty* | (5) *Usefulness* | (6) *Usefulness* |
|---|---|---|---|---|---|---|
| AI Team (Single & Multiple) |  | 0.417*** |  | 0.344** |  | 0.250* |
|  |  | (0.148) |  | (0.150) |  | (0.150) |
| Team age | -0.011 | -0.052 | -0.047 | -0.080 | -0.055 | -0.079 |
|  | (0.073) | (0.073) | (0.069) | (0.070) | (0.069) | (0.071) |
| Team gender | -0.068 | -0.040 | -0.061 | -0.038 | -0.018 | -0.001 |
|  | (0.069) | (0.067) | (0.068) | (0.067) | (0.069) | (0.068) |
| Team education | -0.043 | -0.040 | -0.004 | -0.002 | -0.040 | -0.038 |
|  | (0.056) | (0.057) | (0.058) | (0.058) | (0.054) | (0.055) |
| Team IQ | 0.092 | 0.111 | 0.098 | 0.114 | 0.054 | 0.065 |
|  | (0.074) | (0.073) | (0.074) | (0.075) | (0.079) | (0.079) |
| Team AI use | 0.112 | 0.114 | 0.076 | 0.077 | 0.094 | 0.094 |
|  | (0.074) | (0.072) | (0.075) | (0.074) | (0.074) | (0.074) |
| Team size | 0.126** | 0.092 | 0.137** | 0.108* | 0.097 | 0.076 |
|  | (0.063) | (0.061) | (0.063) | (0.061) | (0.060) | (0.059) |
| Team familiarity | -0.024 | 0.003 | -0.055 | -0.032 | -0.065 | -0.049 |
|  | (0.062) | (0.061) | (0.062) | (0.062) | (0.061) | (0.062) |
| Intercept | 5.955*** | 5.682*** | 5.775*** | 5.549*** | 5.880*** | 5.716*** |
|  | (0.074) | (0.133) | (0.087) | (0.144) | (0.079) | (0.140) |
| Task | Yes | Yes | Yes | Yes | Yes | Yes |
| N | 244 | 244 | 244 | 244 | 244 | 244 |
| BIC | 713.629 | 709.610 | 706.422 | 705.303 | 708.077 | 710.113 |
| Log Pseudolikelihood | -332.077 | -327.319 | -328.474 | -325.166 | -329.301 | -327.571 |
| $R^2$ | 0.050 | 0.087 | 0.051 | 0.076 | 0.035 | 0.049 |

Note: BIC (Bayesian Information Criterion) is a criterion for model selection among a finite set of models; the model with the lowest BIC is preferred. Robust standard errors in parentheses. * $p < 0.1$, ** $p < 0.05$, *** $p < 0.01$.

**Table 2 Generative AI Integration and Team Performance (Experiment I, including single-AI team, multiple-AI team and human-only team conditions)**

|  | (1) *Overall* | (2) *Overall* | (3) *Novelty* | (4) *Novelty* | (5) *Usefulness* | (6) *Usefulness* |
|---|---|---|---|---|---|---|
| *Single-AI team* |  | 0.410** |  | 0.308* |  | 0.198 |
|  |  | (0.161) |  | (0.159) |  | (0.162) |
| *Multiple-AI team* |  | 0.425** |  | 0.385** |  | 0.311* |
|  |  | (0.168) |  | (0.172) |  | (0.172) |
| *Team age* | -0.011 | -0.053 | -0.047 | -0.087 | -0.055 | -0.090 |
|  | (0.073) | (0.075) | (0.069) | (0.074) | (0.069) | (0.073) |
| *Team gender* | -0.068 | -0.041 | -0.061 | -0.040 | -0.018 | -0.004 |
|  | (0.069) | (0.068) | (0.068) | (0.067) | (0.069) | (0.068) |
| *Team education* | -0.043 | -0.040 | -0.004 | 0.002 | -0.040 | -0.034 |
|  | (0.056) | (0.058) | (0.058) | (0.059) | (0.054) | (0.055) |
| *Team IQ* | 0.092 | 0.111 | 0.098 | 0.112 | 0.054 | 0.062 |
|  | (0.074) | (0.074) | (0.074) | (0.075) | (0.079) | (0.080) |
| *Team AI use* | 0.112 | 0.114 | 0.076 | 0.080 | 0.094 | 0.099 |
|  | (0.074) | (0.074) | (0.075) | (0.075) | (0.074) | (0.075) |
| *Team size* | 0.126** | 0.092 | 0.137** | 0.110* | 0.097 | 0.078 |
|  | (0.063) | (0.061) | (0.063) | (0.061) | (0.060) | (0.060) |
| *Team familiarity* | -0.024 | 0.003 | -0.055 | -0.031 | -0.065 | -0.047 |
|  | (0.062) | (0.062) | (0.062) | (0.062) | (0.061) | (0.062) |
| *Intercept* | 5.955*** | 5.681*** | 5.775*** | 5.548*** | 5.880*** | 5.714*** |
|  | (0.074) | (0.133) | (0.087) | (0.145) | (0.079) | (0.140) |
| *Task* | Yes | Yes | Yes | Yes | Yes | Yes |
| *N* | 244 | 244 | 244 | 244 | 244 | 244 |
| *BIC* | 713.629 | 715.098 | 706.422 | 710.537 | 708.077 | 715.041 |
| *Log Pseudolikelihood* | -332.077 | -327.314 | -328.474 | -325.034 | -329.301 | -327.286 |
| $R^2$ | 0.050 | 0.087 | 0.051 | 0.077 | 0.035 | 0.051 |

Note: Robust standard errors in parentheses. * $p < 0.1$, ** $p < 0.05$, *** $p < 0.01$.

'

**Table 3 Generative AI Integration and Team Performance (Experiment I, only AI team)**

|  | (1) *Overall* | (2) *Overall* | (3) *Novelty* | (4) *Novelty* | (5) *Usefulness* | (6) *Usefulness* |
|---|---|---|---|---|---|---|
| Multiple-AI team |  | -0.020 |  | 0.052 |  | 0.092 |
|  |  | (0.138) |  | (0.139) |  | (0.144) |
| Team age | -0.036 | -0.033 | -0.083 | -0.089 | -0.093 | -0.104 |
|  | (0.076) | (0.076) | (0.067) | (0.071) | (0.065) | (0.068) |
| Team gender | -0.042 | -0.041 | -0.036 | -0.037 | -0.006 | -0.010 |
|  | (0.069) | (0.069) | (0.066) | (0.067) | (0.072) | (0.073) |
| Team education | 0.009 | 0.009 | 0.007 | 0.009 | 0.009 | 0.012 |
|  | (0.078) | (0.078) | (0.078) | (0.079) | (0.081) | (0.080) |
| Team IQ | 0.129* | 0.130* | 0.138* | 0.135* | 0.036 | 0.031 |
|  | (0.077) | (0.078) | (0.073) | (0.073) | (0.071) | (0.073) |
| Team AI use | 0.009 | 0.007 | -0.022 | -0.018 | -0.043 | -0.036 |
|  | (0.069) | (0.071) | (0.069) | (0.070) | (0.071) | (0.073) |
| Team size | 0.083 | 0.082 | 0.121* | 0.123* | 0.056 | 0.060 |
|  | (0.067) | (0.068) | (0.063) | (0.064) | (0.066) | (0.068) |
| Team familiarity | -0.005 | -0.006 | -0.011 | -0.010 | -0.095 | -0.093 |
|  | (0.078) | (0.079) | (0.077) | (0.076) | (0.076) | (0.075) |
| Intercept | 6.021*** | 6.031*** | 5.830*** | 5.805*** | 5.878*** | 5.834*** |
|  | (0.080) | (0.101) | (0.093) | (0.105) | (0.084) | (0.101) |
| Task | Yes | Yes | Yes | Yes | Yes | Yes |
| N | 160 | 160 | 160 | 160 | 160 | 160 |
| BIC | 437.732 | 442.785 | 425.010 | 429.927 | 437.808 | 442.430 |
| Log Pseudolikelihood | -196.028 | -196.017 | -189.667 | -189.588 | -196.065 | -195.839 |
| $R^2$ | 0.037 | 0.037 | 0.065 | 0.066 | 0.031 | 0.034 |

Note: Robust standard errors in parentheses. * $p < 0.1$, ** $p < 0.05$, *** $p < 0.01$.

**Table 4 Generative AI Integration and Team Performance (Experiment I, only AI team, Bayesian linear regression)**

| Model | Overall | | | Novelty | | | Usefulness | | |
|---|---|---|---|---|---|---|---|---|---|
| | Mean | Std. Dev. | [95% Cred. Interval] | Mean | Std. Dev. | [95% Cred. Interval] | Mean | Std. Dev. | [95% Cred. Interval] |
| *Team age* | -0.034 | 0.072 | [-0.174, 0.107] | -0.073 | 0.074 | [-0.214, 0.080] | -0.106 | 0.067 | [-0.241, 0.027] |
| *Team gender* | -0.042 | 0.077 | [-0.188, 0.110] | -0.031 | 0.070 | [-0.163, 0.109] | -0.013 | 0.073 | [-0.158, 0.128] |
| *Team education* | -0.016 | 0.092 | [-0.191, 0.173] | 0.011 | 0.088 | [-0.159, 0.188] | 0.023 | 0.089 | [-0.153, 0.195] |
| *Team IQ* | 0.123 | 0.070 | [-0.007, 0.265] | 0.151 | 0.070 | [0.013, 0.297] | 0.052 | 0.062 | [-0.070, 0.177] |
| *Team AI use* | 0.011 | 0.081 | [-0.142, 0.171] | -0.023 | 0.073 | [-0.166, 0.114] | -0.043 | 0.071 | [-0.184, 0.089] |
| *Team size* | 0.079 | 0.073 | [-0.062, 0.224] | 0.121 | 0.066 | [-0.008, 0.249] | 0.051 | 0.066 | [-0.079, 0.183] |
| *Team familiarity* | 0.027 | 0.095 | [-0.175, 0.120] | -0.011 | 0.083 | [-0.168, 0.153] | -0.098 | 0.084 | [-0.261, 0.070] |
| *Intercept* | 6.043 | 0.103 | [5.842, 6.242] | 5.833 | 0.099 | [5.651, 6.027] | 5.899 | 0.085 | [5.734, 6.071] |
| *Multiple-AI* | -0.024 | 0.128 | [-0.273, 0.200] | 0.094 | 0.141 | [-0.170, 0.367] | 0.065 | 0.115 | [-0.154, 0.296] |
| *Team age* | -0.030 | 0.072 | [-0.176, 0.107] | -0.102 | 0.074 | [-0.246, 0.045] | -0.099 | 0.078 | [-0.255, 0.059] |
| *Team gender* | -0.043 | 0.077 | [-0.202, 0.102] | -0.028 | 0.067 | [-0.157, 0.104] | -0.010 | 0.075 | [-0.156, 0.142] |
| *Team education* | 0.005 | 0.086 | [-0.159, 0.178] | 0.016 | 0.082 | [-0.147, 0.178] | -0.005 | 0.082 | [-0.169, 0.156] |
| *Team IQ* | 0.131 | 0.073 | [-0.027, 0.270] | 0.141 | 0.060 | [0.018, 0.256] | 0.020 | 0.065 | [-0.105, 0.142] |
| *Team AI use* | 0.011 | 0.077 | [-0.141, 0.149] | 0.001 | 0.073 | [-0.143, 0.147] | -0.054 | 0.070 | [-0.193, 0.087] |
| *Team size* | 0.093 | 0.069 | [-0.053, 0.219] | 0.133 | 0.068 | [-0.001, 0.261] | 0.069 | 0.067 | [-0.061, 0.210] |
| *Team familiarity* | 0.006 | 0.081 | [-0.156, 0.158] | -0.030 | 0.084 | [-0.200, 0.131] | -0.106 | 0.085 | [-0.270, 0.064] |
| *Intercept* | 6.028 | 0.120 | [5.789, 6.254] | 5.788 | 0.113 | [5.571, 6.011] | 5.832 | 0.112 | [5.619, 6.066] |
| Bayes factor (inverse) | 1355 | | | 1277 | | | 1261 | | |

Note: N=160. We control the task effect in the regression models. MCMC iterations =12500, Burn-in=2500, MCMC sample size=10000.

**Table 5 Heterogeneity effects (Experiment I, only AI team)**

|  | (1) Overall | (2) Novelty | (3) Usefulness | (4) Overall | (5) Novelty | (6) Usefulness | (7) Overall | (8) Novelty | (9) Usefulness |
|---|---|---|---|---|---|---|---|---|---|
| Multiple-AI team*Team IQ | 0.279* | 0.210 | 0.243* | | | | | | |
|  | (0.155) | (0.155) | (0.147) | | | | | | |
| Multiple-AI team*Team size | | | | 0.288** | 0.235* | 0.252* | | | |
|  | | | | (0.138) | (0.139) | (0.142) | | | |
| Multiple-AI team*Team familiarity | | | | | | | 0.248* | 0.275* | 0.278** |
|  | | | | | | | (0.145) | (0.140) | (0.139) |
| Multiple-AI team | 0.015 | 0.078 | 0.122 | -0.064 | 0.017 | 0.054 | 0.020 | 0.096 | 0.136 |
|  | (0.134) | (0.137) | (0.141) | (0.138) | (0.140) | (0.144) | (0.138) | (0.141) | (0.144) |
| Team age | -0.030 | -0.087 | -0.102 | -0.027 | -0.084 | -0.099 | -0.057 | -0.115 | -0.131* |
|  | (0.076) | (0.071) | (0.067) | (0.075) | (0.071) | (0.067) | (0.082) | (0.077) | (0.072) |
| Team gender | -0.041 | -0.037 | -0.010 | -0.039 | -0.036 | -0.008 | -0.048 | -0.045 | -0.017 |
|  | (0.072) | (0.068) | (0.074) | (0.070) | (0.067) | (0.073) | (0.068) | (0.065) | (0.071) |
| Team education | 0.021 | 0.019 | 0.023 | 0.056 | 0.048 | 0.054 | 0.013 | 0.014 | 0.017 |
|  | (0.075) | (0.077) | (0.077) | (0.079) | (0.076) | (0.083) | (0.078) | (0.077) | (0.080) |
| Team IQ | -0.006 | 0.033 | -0.087 | 0.110 | 0.119 | 0.014 | 0.126 | 0.131* | 0.027 |
|  | (0.110) | (0.106) | (0.091) | (0.079) | (0.074) | (0.074) | (0.077) | (0.072) | (0.072) |
| Team AI use | 0.039 | 0.006 | -0.008 | 0.006 | -0.019 | -0.036 | 0.020 | -0.004 | -0.021 |
|  | (0.074) | (0.073) | (0.076) | (0.071) | (0.070) | (0.072) | (0.071) | (0.068) | (0.072) |
| Team size | 0.066 | 0.111* | 0.046 | -0.026 | 0.034 | -0.035 | 0.101 | 0.144** | 0.081 |
|  | (0.068) | (0.064) | (0.068) | (0.080) | (0.076) | (0.081) | (0.068) | (0.063) | (0.067) |
| Team familiarity | -0.003 | -0.008 | -0.091 | 0.005 | -0.001 | -0.084 | -0.118 | -0.135 | -0.219** |
|  | (0.081) | (0.077) | (0.077) | (0.082) | (0.080) | (0.077) | (0.096) | (0.093) | (0.088) |
| Intercept | 6.009*** | 5.788*** | 5.815*** | 6.042*** | 5.814*** | 5.843*** | 6.015*** | 5.787*** | 5.816*** |
|  | (0.100) | (0.104) | (0.102) | (0.101) | (0.105) | (0.102) | (0.100) | (0.105) | (0.100) |
| Task | Yes | Yes | Yes | Yes | Yes | Yes | Yes | Yes | Yes |
| N | 160 | 160 | 160 | 160 | 160 | 160 | 160 | 160 | 160 |
| BIC | 443.698 | 432.461 | 444.342 | 443.689 | 431.991 | 444.301 | 445.658 | 432.058 | 444.727 |
| Log Pseudolikelihood | -193.935 | -188.317 | -194.257 | -193.931 | -188.082 | -194.237 | -194.915 | -188.116 | -194.450 |
| $R^2$ | 0.062 | 0.081 | 0.053 | 0.062 | 0.084 | 0.053 | 0.050 | 0.083 | 0.050 |

Note: Robust standard errors in parentheses. * $p < 0.1$, ** $p < 0.05$, *** $p < 0.01$.

**Table 6 Team Performance VS Individual-AI Pair Performance (Experiment II)**

|  | (1) Overall | (2) Overall | (3) Novelty | (4) Novelty | (5) Usefulness | (6) Usefulness |
|---|---|---|---|---|---|---|
| Individual-AI pair |  | -0.378** |  | -0.410*** |  | -0.438*** |
|  |  | (0.151) |  | (0.153) |  | (0.152) |
| AI team (Single & Multiple) |  | 0.463*** |  | 0.411*** |  | 0.295** |
|  |  | (0.144) |  | (0.145) |  | (0.145) |
| Age | 0.033 | -0.067 | -0.001 | -0.099 | -0.005 | -0.092 |
|  | (0.084) | (0.080) | (0.080) | (0.076) | (0.083) | (0.081) |
| Gender | -0.047 | -0.035 | -0.045 | -0.034 | -0.045 | -0.038 |
|  | (0.063) | (0.062) | (0.062) | (0.062) | (0.063) | (0.063) |
| Education | 0.036 | 0.015 | 0.072 | 0.052 | 0.059 | 0.039 |
|  | (0.073) | (0.067) | (0.072) | (0.066) | (0.074) | (0.069) |
| IQ | 0.119** | 0.159*** | 0.126** | 0.165*** | 0.089 | 0.123** |
|  | (0.060) | (0.059) | (0.059) | (0.058) | (0.062) | (0.062) |
| AI use | 0.179*** | 0.153*** | 0.161*** | 0.135** | 0.178*** | 0.154*** |
|  | (0.058) | (0.057) | (0.058) | (0.057) | (0.059) | (0.059) |
| Intercept | 5.550*** | 5.609*** | 5.372*** | 5.464*** | 5.515*** | 5.658*** |
|  | (0.074) | (0.131) | (0.078) | (0.138) | (0.077) | (0.134) |
| Task | Yes | Yes | Yes | Yes | Yes | Yes |
| N | 522 | 522 | 522 | 522 | 522 | 522 |
| BIC | 1755.059 | 1724.500 | 1757.490 | 1728.924 | 1770.093 | 1750.019 |
| Log Pseudolikelihood | -855.628 | -834.090 | -856.843 | -836.303 | -863.145 | -846.850 |
| $R^2$ | 0.034 | 0.110 | 0.032 | 0.105 | 0.029 | 0.087 |

Note: Robust standard errors in parentheses. * $p < 0.1$, ** $p < 0.05$, *** $p < 0.01$.

**Table 7 Task Completion Time for Different Conditions**

|  | (1) | (2) | (3) |
|---|---|---|---|
|  |  | *Task Completion Time* |  |
| *AI team* | 0.037 |  | 0.049 |
|  | (0.068) |  | (0.071) |
| *Multiple-AI team* |  | 0.128* |  |
|  |  | (0.068) |  |
| *Individual-AI pair* |  |  | -0.659*** |
|  |  |  | (0.069) |
| *Team age* | -0.004 | -0.030 | -0.025 |
|  | (0.032) | (0.036) | (0.035) |
| *Team gender* | 0.075** | 0.071* | 0.017 |
|  | (0.029) | (0.039) | (0.024) |
| *Team education* | 0.020 | 0.116** | 0.011 |
|  | (0.039) | (0.056) | (0.036) |
| *Team IQ* | 0.002 | 0.015 | 0.036 |
|  | (0.029) | (0.036) | (0.025) |
| *Team AI use* | 0.059** | 0.030 | 0.076*** |
|  | (0.030) | (0.038) | (0.025) |
| *Team size* | 0.007 | 0.022 | -- |
|  | (0.029) | (0.035) | -- |
| *Team familiarity* | -0.006 | 0.017 | -- |
|  | (0.033) | (0.046) | -- |
| *Intercept* | 3.496*** | 3.497*** | 3.673*** |
|  | (0.063) | (0.065) | (0.064) |
| Task | Yes | Yes | Yes |
| N | 235 | 156 | 513 |
| BIC | 327.348 | 210.381 | 871.646 |
| Log Pseudolikelihood | -136.376 | -79.941 | -407.742 |
| $R^2$ | 0.193 | 0.218 | 0.302 |

Note: Robust standard errors in parentheses. * $p < 0.1$, ** $p < 0.05$, *** $p < 0.01$.

**Table 8 AI Team (Single & Multiple) and Task Completion Time (Experiment I, only AI team, Bayesian linear regression)**

| Model | Task Completion Time | | |
|---|---|---|---|
| | Mean | Std. Dev. | [95% Cred. Interval] |
| *Team age* | 0.001 | 0.031 | [-0.058, 0.061] |
| *Team gender* | 0.071 | 0.027 | [0.020, 0.124] |
| *Team education* | 0.027 | 0.029 | [-0.028, 0.083] |
| *Team IQ* | 0.000 | 0.026 | [-0.051, 0.050] |
| *Team AI use* | 0.053 | 0.028 | [-0.004, 0.106] |
| *Team size* | 0.010 | 0.029 | [-0.046, 0.068] |
| *Team familiarity* | -0.011 | 0.028 | [-0.067, 0.045] |
| *Intercept* | 3.513 | 0.041 | [3.434, 3.595] |
| *AI Team* (*Single & Multiple*) | 0.057 | 0.064 | [-0.083, 0.164] |
| *Team age* | -0.005 | 0.032 | [-0.071, 0.060] |
| *Team gender* | 0.073 | 0.030 | [0.018, 0.130] |
| *Team education* | 0.023 | 0.032 | [-0.046, 0.086] |
| *Team IQ* | -0.006 | 0.024 | [-0.056, 0.039] |
| *Team AI use* | 0.078 | 0.029 | [0.021, 0.134] |
| *Team size* | 0.009 | 0.030 | [-0.050, 0.071] |
| *Team familiarity* | -0.002 | 0.027 | [-0.053, 0.051] |
| *Intercept* | 3.507 | 0.061 | [3.387, 3.635] |
| *Bayes factor (inverse)* | 1333 | | |

Note: N=235. We control the task effect in the regression models. MCMC iterations =12500, Burn-in=2500, MCMC sample size=10000.

**Table 9 Team Performance VS Individual-AI Pair Performance (Experiment II, control task completion time)**

|  | (1) Overall | (2) Overall | (3) Novelty | (4) Novelty | (5) Usefulness | (6) Usefulness |
|---|---|---|---|---|---|---|
| *Individual-AI pair* |  | -0.153 |  | -0.173 |  | -0.188 |
|  |  | (0.163) |  | (0.164) |  | (0.163) |
| *AI team (Single & Multiple)* |  | 0.422*** |  | 0.372** |  | 0.251* |
|  |  | (0.144) |  | (0.147) |  | (0.145) |
| *Age* | -0.023 | -0.078 | -0.057 | -0.109 | -0.063 | -0.104 |
|  | (0.078) | (0.077) | (0.073) | (0.073) | (0.078) | (0.078) |
| *Gender* | -0.058 | -0.046 | -0.054 | -0.044 | -0.055 | -0.048 |
|  | (0.061) | (0.062) | (0.061) | (0.061) | (0.062) | (0.062) |
| *Education* | 0.034 | 0.025 | 0.070 | 0.060 | 0.059 | 0.051 |
|  | (0.067) | (0.065) | (0.064) | (0.063) | (0.068) | (0.067) |
| *IQ* | 0.120** | 0.147** | 0.126** | 0.151*** | 0.089 | 0.109* |
|  | (0.059) | (0.059) | (0.057) | (0.058) | (0.062) | (0.062) |
| *AI use* | 0.128** | 0.129** | 0.110* | 0.110** | 0.127** | 0.127** |
|  | (0.057) | (0.056) | (0.057) | (0.056) | (0.059) | (0.058) |
| *Task completion time* | 0.364*** | 0.258*** | 0.364*** | 0.261*** | 0.374*** | 0.287*** |
|  | (0.051) | (0.054) | (0.052) | (0.054) | (0.052) | (0.055) |
| *Intercept* | 5.579*** | 5.527*** | 5.404*** | 5.378*** | 5.548*** | 5.568*** |
|  | (0.070) | (0.132) | (0.074) | (0.139) | (0.074) | (0.134) |
| *Task* | Yes | Yes | Yes | Yes | Yes | Yes |
| *N* | 513 | 513 | 513 | 513 | 513 | 513 |
| *BIC* | 1687.577 | 1682.780 | 1689.270 | 1686.586 | 1701.462 | 1704.678 |
| *Log Pseudolikelihood* | -818.827 | -810.189 | -819.674 | -812.092 | -825.770 | -821.138 |
| *R2* | 0.113 | 0.142 | 0.111 | 0.137 | 0.111 | 0.127 |

Note: Robust standard errors in parentheses. * $p < 0.1$, ** $p < 0.05$, *** $p < 0.01$.

**Table 10 Team Performance VS Individual-AI Pair Performance (Experiment II, Control task completion time, Bayesian linear regression)**

| Model | Overall | | | Novelty | | | Usefulness | | |
|---|---|---|---|---|---|---|---|---|---|
| | Mean | Std. Dev. | [95% Cred. Interval] | Mean | Std. Dev. | [95% Cred. Interval] | Mean | Std. Dev. | [95% Cred. Interval] |
| *AI team (Single & Multiple)* | 0.489 | 0.132 | [0.244, 0.754] | 0.484 | 0.143 | [0.191, 0.765] | 0.357 | 0.130 | [0.100, 0.615] |
| *Age* | -0.076 | 0.064 | [-0.198, 0.062] | -0.109 | 0.063 | [-0.242, 0.009] | -0.105 | 0.061 | [-0.224, 0.022] |
| *Gender* | -0.043 | 0.054 | [-0.148, 0.068] | -0.052 | 0.053 | [-0.161, 0.052] | -0.044 | 0.054 | [-0.149, 0.066] |
| *Education* | 0.039 | 0.058 | [-0.078, 0.145] | 0.060 | 0.057 | [-0.057, 0.169] | 0.060 | 0.059 | [-0.054, 0.176] |
| *IQ* | 0.146 | 0.054 | [0.036, 0.250] | 0.140 | 0.053 | [0.030, 0.245] | 0.099 | 0.052 | [-0.007, 0.206] |
| *AI use* | 0.123 | 0.054 | [0.026, 0.232] | 0.112 | 0.057 | [0.002, 0.223] | 0.138 | 0.048 | [0.050, 0.231] |
| *Task completion time* | 0.298 | 0.062 | [0.179, 0.417] | 0.288 | 0.063 | [0.163, 0.416] | 0.314 | 0.055 | [0.207, 0.416] |
| *Intercept* | 5.425 | 0.085 | [5.252, 5.582] | 5.246 | 0.090 | [5.089, 5.446] | 5.446 | 0.078 | [5.287, 5.591] |
| *Individual* | -0.196 | 0.171 | [-0.541, 0.133] | -0.175 | 0.163 | [-0.490, 0.142] | -0.200 | 0.154 | [-0.504, 0.103] |
| *AI team (Single & Multiple)* | 0.387 | 0.162 | [0.067, 0.701] | 0.347 | 0.158 | [0.041, 0.673] | 0.226 | 0.169 | [-0.096, 0.537] |
| *Age* | -0.072 | 0.062 | [-0.188, 0.058] | -0.121 | 0.059 | [-0.238, -0.012] | -0.098 | 0.057 | [-0.211, 0.019] |
| *Gender* | -0.037 | 0.056 | [-0.148, 0.072] | -0.049 | 0.054 | [-0.159, 0.055] | -0.026 | 0.046 | [-0.117, 0.059] |
| *Education* | 0.018 | 0.057 | [-0.094, 0.129] | 0.073 | 0.055 | [-0.034, 0.172] | 0.035 | 0.053 | [-0.065, 0.140] |
| *IQ* | 0.148 | 0.057 | [0.035, 0.262] | 0.147 | 0.051 | [0.041, 0.247] | 0.110 | 0.050 | [0.013, 0.206] |
| *AI use* | 0.124 | 0.059 | [0.007, 0.241] | 0.113 | 0.052 | [0.009, 0.216] | 0.128 | 0.060 | [0.011, 0.247] |
| *Task completion time* | 0.256 | 0.062 | [0.138, 0.379] | 0.263 | 0.060 | [0.139, 0.378] | 0.274 | 0.057 | [0.159, 0.378] |
| *Intercept* | 5.559 | 0.141 | [5.294, 5.819] | 5.418 | 0.140 | [5.141, 5.688] | 5.594 | 0.133 | [5.340, 5.849] |
| Bayes factor (inverse) | 263 | | | 714 | | | 909 | | |

Note: N=513. We control the task effect in the regression models. MCMC iterations =12500, Burn-in=2500, MCMC sample size=10000.

**Table 11 Human-Only Team Performance VS Individual-AI Pair Performance (Experiment II, CEM Sample)**

|  | (1) *Overall* | (2) *Overall* | (3) *Novelty* | (4) *Novelty* | (5) *Usefulness* | (6) *Usefulness* |
|---|---|---|---|---|---|---|
| *Individual-AI pair* |  | -0.097 |  | -0.126 |  | -0.184 |
|  |  | (0.238) |  | (0.252) |  | (0.248) |
| *Age* | -0.023 | -0.025 | -0.006 | -0.008 | -0.021 | -0.023 |
|  | (0.119) | (0.119) | (0.133) | (0.134) | (0.121) | (0.121) |
| *Gender* | 0.016 | 0.010 | 0.022 | 0.015 | -0.017 | -0.027 |
|  | (0.128) | (0.133) | (0.132) | (0.137) | (0.131) | (0.136) |
| *Education* | -0.173 | -0.177 | -0.117 | -0.124 | -0.109 | -0.118 |
|  | (0.118) | (0.121) | (0.125) | (0.128) | (0.120) | (0.123) |
| *IQ* | 0.225 | 0.226* | 0.200 | 0.203 | 0.220 | 0.224 |
|  | (0.137) | (0.136) | (0.150) | (0.149) | (0.153) | (0.150) |
| *AI use* | 0.234** | 0.232** | 0.239** | 0.236* | 0.309*** | 0.306** |
|  | (0.107) | (0.109) | (0.117) | (0.119) | (0.115) | (0.118) |
| *Intercept* | 5.711*** | 5.760*** | 5.547*** | 5.610*** | 5.725*** | 5.819*** |
|  | (0.129) | (0.145) | (0.142) | (0.165) | (0.142) | (0.157) |
| *Task* | Yes | Yes | Yes | Yes | Yes | Yes |
| *N* | 110 | 110 | 110 | 110 | 110 | 110 |
| *BIC* | 371.299 | 375.802 | 382.650 | 387.050 | 380.559 | 384.603 |
| *Log Pseudolikelihood* | -169.198 | -169.099 | -174.873 | -174.723 | -173.828 | -173.500 |
| $R^2$ | 0.098 | 0.100 | 0.072 | 0.075 | 0.104 | 0.109 |

Note: Controlling for task completion time based on the data from Table 6, we obtained the same results. Robust standard errors in parentheses. * $p < 0.1$, ** $p < 0.05$, *** $p < 0.01$

**Table 12 AI Team Performance VS Individual-AI Pair Performance (Experiment II, CEM Sample)**

|  | (1) *Overall* | (2) *Overall* | (3) *Novelty* | (4) *Novelty* | (5) *Usefulness* | (6) *Usefulness* |
|---|---|---|---|---|---|---|
| *Individual-AI pair* |  | -0.706*** |  | -0.689*** |  | -0.577*** |
|  |  | (0.208) |  | (0.211) |  | (0.218) |
| *Age* | 0.016 | -0.076 | -0.039 | -0.129 | -0.038 | -0.113 |
|  | (0.147) | (0.130) | (0.135) | (0.117) | (0.143) | (0.130) |
| *Gender* | -0.134 | -0.095 | -0.107 | -0.069 | -0.103 | -0.071 |
|  | (0.144) | (0.141) | (0.140) | (0.137) | (0.142) | (0.140) |
| *Education* | 0.064 | 0.035 | 0.068 | 0.040 | 0.046 | 0.022 |
|  | (0.119) | (0.110) | (0.117) | (0.108) | (0.122) | (0.116) |
| *IQ* | 0.316*** | 0.372*** | 0.323*** | 0.378*** | 0.262** | 0.309*** |
|  | (0.114) | (0.114) | (0.105) | (0.107) | (0.117) | (0.118) |
| *AI use* | 0.165* | 0.161* | 0.121 | 0.117 | 0.180* | 0.177** |
|  | (0.095) | (0.089) | (0.097) | (0.091) | (0.093) | (0.089) |
| *Intercept* | 5.599*** | 5.958*** | 5.461*** | 5.812*** | 5.568*** | 5.862*** |
|  | (0.123) | (0.124) | (0.128) | (0.134) | (0.125) | (0.124) |
| *Task* | Yes | Yes | Yes | Yes | Yes | Yes |
| *N* | 154 | 154 | 154 | 154 | 154 | 154 |
| *BIC* | 535.402 | 528.875 | 535.708 | 529.759 | 538.573 | 536.132 |
| *Log Pseudolikelihood* | -250.072 | -244.290 | -250.224 | -244.732 | -251.657 | -247.918 |
| *R²* | 0.086 | 0.152 | 0.083 | 0.146 | 0.071 | 0.115 |

Note: Robust standard errors in parentheses. * $p < 0.1$, ** $p < 0.05$, *** $p < 0.01$.

**Table 13 Mechanism Analysis**

|  | (1) Overall | (2) Novelty | (3) Usefulness | (4) Overall | (5) Novelty | (6) Usefulness | (7) Overall | (8) Novelty | (9) Usefulness |
|---|---|---|---|---|---|---|---|---|---|
| Total prompt length | 0.144* | 0.129** | 0.146* | | | | | | |
|  | (0.082) | (0.062) | (0.078) | | | | | | |
| Total interaction round | | | | 0.169*** | 0.210*** | 0.200*** | | | |
|  | | | | (0.056) | (0.063) | (0.060) | | | |
| Human input | | | | | | | 0.202*** | 0.154*** | 0.174*** |
|  | | | | | | | (0.066) | (0.058) | (0.065) |
| Team age | -0.047 | -0.094 | -0.106 | -0.043 | -0.096 | -0.104 | -0.063 | -0.104 | -0.117* |
|  | (0.077) | (0.068) | (0.066) | (0.077) | (0.069) | (0.067) | (0.076) | (0.066) | (0.064) |
| Team gender | -0.052 | -0.041 | -0.014 | -0.079 | -0.075 | -0.046 | -0.039 | -0.031 | -0.002 |
|  | (0.067) | (0.064) | (0.070) | (0.069) | (0.065) | (0.072) | (0.067) | (0.065) | (0.070) |
| Team education | 0.050 | 0.041 | 0.048 | 0.034 | 0.034 | 0.036 | 0.040 | 0.030 | 0.035 |
|  | (0.076) | (0.078) | (0.080) | (0.073) | (0.073) | (0.076) | (0.079) | (0.079) | (0.081) |
| Team IQ | 0.154** | 0.148** | 0.051 | 0.157** | 0.145** | 0.051 | 0.175** | 0.166** | 0.072 |
|  | (0.070) | (0.069) | (0.068) | (0.068) | (0.066) | (0.066) | (0.068) | (0.070) | (0.067) |
| Team AI use | 0.023 | -0.009 | -0.028 | 0.028 | 0.000 | -0.021 | 0.019 | -0.014 | -0.034 |
|  | (0.069) | (0.068) | (0.070) | (0.067) | (0.066) | (0.068) | (0.070) | (0.069) | (0.072) |
| Team size | 0.057 | 0.099 | 0.031 | 0.066 | 0.103* | 0.038 | 0.058 | 0.102 | 0.035 |
|  | (0.068) | (0.064) | (0.067) | (0.066) | (0.061) | (0.065) | (0.067) | (0.062) | (0.065) |
| Team familiarity | 0.001 | -0.009 | -0.091 | -0.003 | -0.016 | -0.097 | 0.012 | 0.000 | -0.081 |
|  | (0.077) | (0.075) | (0.074) | (0.075) | (0.072) | (0.072) | (0.077) | (0.078) | (0.073) |
| Intercept | 6.025*** | 5.836*** | 5.880*** | 5.981*** | 5.781*** | 5.829*** | 6.017*** | 5.831*** | 5.874*** |
|  | (0.081) | (0.093) | (0.085) | (0.083) | (0.091) | (0.085) | (0.080) | (0.094) | (0.085) |
| Task | Yes | Yes | Yes | Yes | Yes | Yes | Yes | Yes | Yes |
| N | 158 | 158 | 158 | 158 | 158 | 158 | 158 | 158 | 158 |
| BIC | 425.741 | 416.353 | 428.906 | 424.255 | 410.090 | 425.125 | 420.772 | 414.352 | 426.635 |
| Log Pseudolikelihood | -187.557 | -182.863 | -189.140 | -186.815 | -179.732 | -187.250 | -185.073 | -181.863 | -188.005 |
| $R^2$ | 0.082 | 0.100 | 0.063 | 0.091 | 0.135 | 0.085 | 0.111 | 0.111 | 0.076 |

Note: Robust standard errors in parentheses. * $p < 0.1$, ** $p < 0.05$, *** $p < 0.01$.

**Table 14 Robustness of Mechanism Analysis**

|  | (1) Overall | (2) Novelty | (3) Usefulness | (4) Overall | (5) Novelty | (6) Usefulness | (7) Overall | (8) Novelty | (9) Usefulness |
|---|---|---|---|---|---|---|---|---|---|
| *Average prompt length* | 0.202*** | 0.168*** | 0.161*** |  |  |  |  |  |  |
|  | (0.051) | (0.042) | (0.047) |  |  |  |  |  |  |
| *Average interaction round* |  |  |  | 0.187*** | 0.200*** | 0.162** |  |  |  |
|  |  |  |  | (0.065) | (0.071) | (0.068) |  |  |  |
| *Human input_Top50* |  |  |  |  |  |  | 0.228*** | 0.191*** | 0.200*** |
|  |  |  |  |  |  |  | (0.063) | (0.055) | (0.064) |
| *Team age* | -0.049 | -0.095 | -0.103 | -0.024 | -0.073 | -0.083 | -0.069 | -0.112* | -0.122* |
|  | (0.076) | (0.067) | (0.065) | (0.076) | (0.067) | (0.064) | (0.074) | (0.064) | (0.063) |
| *Team gender* | -0.050 | -0.039 | -0.012 | -0.069 | -0.059 | -0.029 | -0.050 | -0.040 | -0.012 |
|  | (0.063) | (0.062) | (0.068) | (0.066) | (0.064) | (0.070) | (0.066) | (0.064) | (0.070) |
| *Team education* | 0.042 | 0.033 | 0.035 | 0.016 | 0.011 | 0.014 | 0.028 | 0.022 | 0.025 |
|  | (0.076) | (0.077) | (0.081) | (0.073) | (0.074) | (0.079) | (0.078) | (0.077) | (0.080) |
| *Team IQ* | 0.159** | 0.153** | 0.059 | 0.171** | 0.162** | 0.068 | 0.149** | 0.145** | 0.049 |
|  | (0.068) | (0.068) | (0.066) | (0.066) | (0.066) | (0.065) | (0.072) | (0.070) | (0.070) |
| *Team AI use* | 0.018 | -0.015 | -0.036 | 0.008 | -0.023 | -0.043 | 0.017 | -0.015 | -0.035 |
|  | (0.069) | (0.069) | (0.071) | (0.069) | (0.068) | (0.071) | (0.071) | (0.069) | (0.072) |
| *Team size* | 0.076 | 0.116* | 0.051 | 0.087 | 0.127** | 0.060 | 0.069 | 0.110* | 0.045 |
|  | (0.065) | (0.062) | (0.065) | (0.064) | (0.060) | (0.065) | (0.064) | (0.060) | (0.064) |
| *Team familiarity* | 0.013 | 0.002 | -0.080 | -0.007 | -0.019 | -0.098 | 0.011 | -0.000 | -0.082 |
|  | (0.076) | (0.075) | (0.074) | (0.075) | (0.073) | (0.073) | (0.076) | (0.076) | (0.074) |
| *Intercept* | 6.035*** | 5.845*** | 5.889*** | 5.975*** | 5.782*** | 5.838*** | 6.011*** | 5.825*** | 5.869*** |
|  | (0.080) | (0.093) | (0.084) | (0.086) | (0.097) | (0.090) | (0.081) | (0.093) | (0.085) |
| Task | Yes | Yes | Yes | Yes | Yes | Yes | Yes | Yes | Yes |
| N | 158 | 158 | 158 | 158 | 158 | 158 | 158 | 158 | 158 |
| BIC | 420.374 | 412.923 | 427.395 | 422.312 | 410.351 | 427.667 | 417.773 | 410.979 | 424.238 |
| Log Pseudolikelihood | -184.874 | -181.149 | -188.384 | -185.843 | -179.862 | -188.520 | -183.574 | -180.177 | -186.806 |
| $R^2$ | 0.113 | 0.119 | 0.072 | 0.102 | 0.133 | 0.070 | 0.127 | 0.130 | 0.090 |

Note: Robust standard errors in parentheses. * $p < 0.1$, ** $p < 0.05$, *** $p < 0.01$.

**Table 15 Interaction Pattern**

|  | (1) Overall | (2) Novelty | (3) Usefulness | (4) Overall | (5) Novelty | (6) Usefulness |
|---|---|---|---|---|---|---|
| *Gini of prompt length* | 0.187** | 0.249** | 0.196** | | | |
|  | (0.089) | (0.094) | (0.094) | | | |
| *Gini of interaction round* | | | | 0.057 | 0.054 | -0.040 |
|  | | | | (0.098) | (0.106) | (0.107) |
| *Team age* | -0.146 | -0.208** | -0.235** | -0.139 | -0.210* | -0.276** |
|  | (0.093) | (0.094) | (0.095) | (0.107) | (0.109) | (0.111) |
| *Team gender* | -0.070 | -0.074 | -0.026 | -0.054 | -0.057 | -0.029 |
|  | (0.100) | (0.103) | (0.112) | (0.111) | (0.116) | (0.123) |
| *Team education* | 0.192 | 0.204 | 0.246 | 0.162 | 0.162 | 0.204 |
|  | (0.164) | (0.179) | (0.174) | (0.165) | (0.179) | (0.177) |
| *Team IQ* | 0.234** | 0.246** | 0.133 | 0.198* | 0.194 | 0.082 |
|  | (0.099) | (0.108) | (0.112) | (0.105) | (0.117) | (0.117) |
| *Team AI use* | 0.048 | -0.033 | -0.036 | 0.048 | -0.030 | -0.028 |
|  | (0.107) | (0.109) | (0.107) | (0.108) | (0.111) | (0.109) |
| *Team size* | 0.295** | 0.338** | 0.275** | 0.256** | 0.288** | 0.242* |
|  | (0.127) | (0.130) | (0.133) | (0.128) | (0.131) | (0.137) |
| *Team familiarity* | 0.150 | 0.160 | 0.070 | 0.141 | 0.155 | 0.089 |
|  | (0.112) | (0.108) | (0.124) | (0.121) | (0.123) | (0.137) |
| *Intercept* | 6.056*** | 6.003*** | 5.969*** | 6.042*** | 5.986*** | 5.964*** |
|  | (0.126) | (0.140) | (0.138) | (0.133) | (0.158) | (0.147) |
| *Task* | Yes | Yes | Yes | Yes | Yes | Yes |
| *N* | 74 | 74 | 74 | 74 | 74 | 74 |
| *BIC* | 210.045 | 213.206 | 219.833 | 213.732 | 219.628 | 223.568 |
| *Log Pseudolikelihood* | -83.502 | -85.083 | -88.396 | -85.346 | -88.293 | -90.264 |
| $R^2$ | 0.189 | 0.267 | 0.174 | 0.148 | 0.201 | 0.131 |

Note: Robust standard errors in parentheses. * $p < 0.1$, ** $p < 0.05$, *** $p < 0.01$.

**Table 16 Generative AI Integration and Team Process (Experiment I, including human-only team condition)**

|  | (1) | (2) | (3) | (4) | (5) | (6) | (7) | (8) |
|---|---|---|---|---|---|---|---|---|
|  | *Team Coordination* | | *Team Potency* | | *Team Information Elaboration* | | *Team Satisfaction* | |
| *AI team* (*Single & Multiple*) |  | 0.010 |  | 0.313** |  | 0.105 |  | 0.266* |
|  |  | (0.124) |  | (0.149) |  | (0.126) |  | (0.149) |
| *Team age* | -0.007 | -0.008 | 0.115** | 0.085 | 0.054 | 0.043 | 0.035 | 0.009 |
|  | (0.059) | (0.062) | (0.052) | (0.054) | (0.051) | (0.053) | (0.071) | (0.072) |
| *Team gender* | -0.035 | -0.034 | -0.021 | 0.000 | -0.012 | -0.004 | -0.040 | -0.022 |
|  | (0.059) | (0.059) | (0.063) | (0.060) | (0.059) | (0.061) | (0.069) | (0.067) |
| *Team education* | 0.039 | 0.039 | 0.012 | 0.014 | -0.027 | -0.026 | 0.040 | 0.042 |
|  | (0.052) | (0.053) | (0.062) | (0.062) | (0.051) | (0.051) | (0.063) | (0.063) |
| *Team IQ* | 0.022 | 0.022 | -0.053 | -0.038 | 0.012 | 0.017 | -0.043 | -0.030 |
|  | (0.067) | (0.067) | (0.064) | (0.065) | (0.060) | (0.060) | (0.069) | (0.069) |
| *Team AI use* | -0.044 | -0.044 | -0.094 | -0.093 | -0.082 | -0.081 | -0.109 | -0.108 |
|  | (0.061) | (0.061) | (0.066) | (0.064) | (0.061) | (0.061) | (0.072) | (0.072) |
| *Team size* | -0.058 | -0.059 | -0.114** | -0.140*** | -0.048 | -0.056 | -0.073 | -0.095* |
|  | (0.048) | (0.048) | (0.055) | (0.051) | (0.052) | (0.050) | (0.059) | (0.056) |
| *Team familiarity* | 0.107** | 0.107** | 0.141*** | 0.161*** | 0.057 | 0.064 | 0.148*** | 0.166*** |
|  | (0.044) | (0.045) | (0.052) | (0.055) | (0.051) | (0.052) | (0.053) | (0.053) |
| *Intercept* | 6.100*** | 6.093*** | 5.690*** | 5.485*** | 5.999*** | 5.930*** | 5.839*** | 5.665*** |
|  | (0.050) | (0.104) | (0.058) | (0.127) | (0.051) | (0.106) | (0.060) | (0.133) |
| *N* | 122 | 122 | 122 | 122 | 122 | 122 | 122 | 122 |
| BIC | 229.130 | 233.926 | 268.196 | 266.744 | 238.378 | 242.312 | 277.981 | 278.651 |
| *Log Pseudolikelihood* | -95.349 | -95.345 | -114.882 | -111.754 | -99.973 | -99.538 | -119.774 | -117.708 |
| $R^2$ | 0.064 | 0.064 | 0.125 | 0.168 | 0.046 | 0.053 | 0.095 | 0.125 |

Note: Robust standard errors in parentheses. * $p < 0.1$, ** $p < 0.05$, *** $p < 0.01$.

'

**Table 17 Generative AI Integration and Team Process (Experiment I, only AI team)**

|  | (1) | (2) | (3) | (4) | (5) | (6) | (7) | (8) |
|---|---|---|---|---|---|---|---|---|
|  | \multicolumn{2}{c}{*Team Coordination*} | \multicolumn{2}{c}{*Team Potency*} | \multicolumn{2}{c}{*Team Information Elaboration*} | \multicolumn{2}{c}{*Team Satisfaction*} |
| *Multiple-AI team* |  | -0.021 |  | -0.074 |  | -0.003 |  | 0.031 |
|  |  | (0.134) |  | (0.128) |  | (0.125) |  | (0.141) |
| *Team age* | 0.081 | 0.083 | 0.165*** | 0.174*** | 0.095* | 0.095* | 0.100 | 0.096 |
|  | (0.064) | (0.071) | (0.051) | (0.051) | (0.054) | (0.053) | (0.078) | (0.085) |
| *Team gender* | -0.050 | -0.049 | -0.032 | -0.029 | -0.042 | -0.042 | -0.029 | -0.031 |
|  | (0.067) | (0.068) | (0.058) | (0.058) | (0.070) | (0.070) | (0.068) | (0.068) |
| *Team education* | 0.033 | 0.032 | 0.016 | 0.013 | -0.018 | -0.018 | 0.014 | 0.015 |
|  | (0.078) | (0.080) | (0.088) | (0.088) | (0.076) | (0.077) | (0.087) | (0.091) |
| *Team IQ* | 0.061 | 0.062 | -0.029 | -0.025 | 0.003 | 0.003 | 0.007 | 0.006 |
|  | (0.082) | (0.083) | (0.070) | (0.069) | (0.081) | (0.079) | (0.082) | (0.082) |
| *Team AI use* | -0.074 | -0.075 | -0.118 | -0.124 | -0.095 | -0.095 | -0.100 | -0.098 |
|  | (0.074) | (0.078) | (0.078) | (0.082) | (0.076) | (0.081) | (0.082) | (0.084) |
| *Team size* | -0.074 | -0.075 | -0.148** | -0.151** | -0.033 | -0.033 | -0.121* | -0.119* |
|  | (0.057) | (0.059) | (0.058) | (0.061) | (0.057) | (0.058) | (0.066) | (0.068) |
| *Team familiarity* | 0.097* | 0.097* | 0.104 | 0.103 | 0.036 | 0.036 | 0.147** | 0.148** |
|  | (0.051) | (0.050) | (0.075) | (0.075) | (0.058) | (0.059) | (0.062) | (0.061) |
| *Intercept* | 6.092*** | 6.102*** | 5.778*** | 5.813*** | 6.014*** | 6.016*** | 5.923*** | 5.908*** |
|  | (0.060) | (0.090) | (0.065) | (0.084) | (0.061) | (0.079) | (0.063) | (0.099) |
| N | 80 | 80 | 80 | 80 | 80 | 80 | 80 | 80 |
| BIC | 145.134 | 149.480 | 155.526 | 159.528 | 152.329 | 156.710 | 161.707 | 166.029 |
| Log Pseudolikelihood | -55.039 | -55.021 | -60.235 | -60.045 | -58.636 | -58.636 | -63.326 | -63.295 |
| $R^2$ | 0.103 | 0.104 | 0.190 | 0.194 | 0.068 | 0.068 | 0.133 | 0.134 |

Note: Robust standard errors in parentheses. * $p < 0.1$, ** $p < 0.05$, *** $p < 0.01$.

**Figure 1 Relative Distribution of Individual-AI Pair and Human-Only Team Conditions**

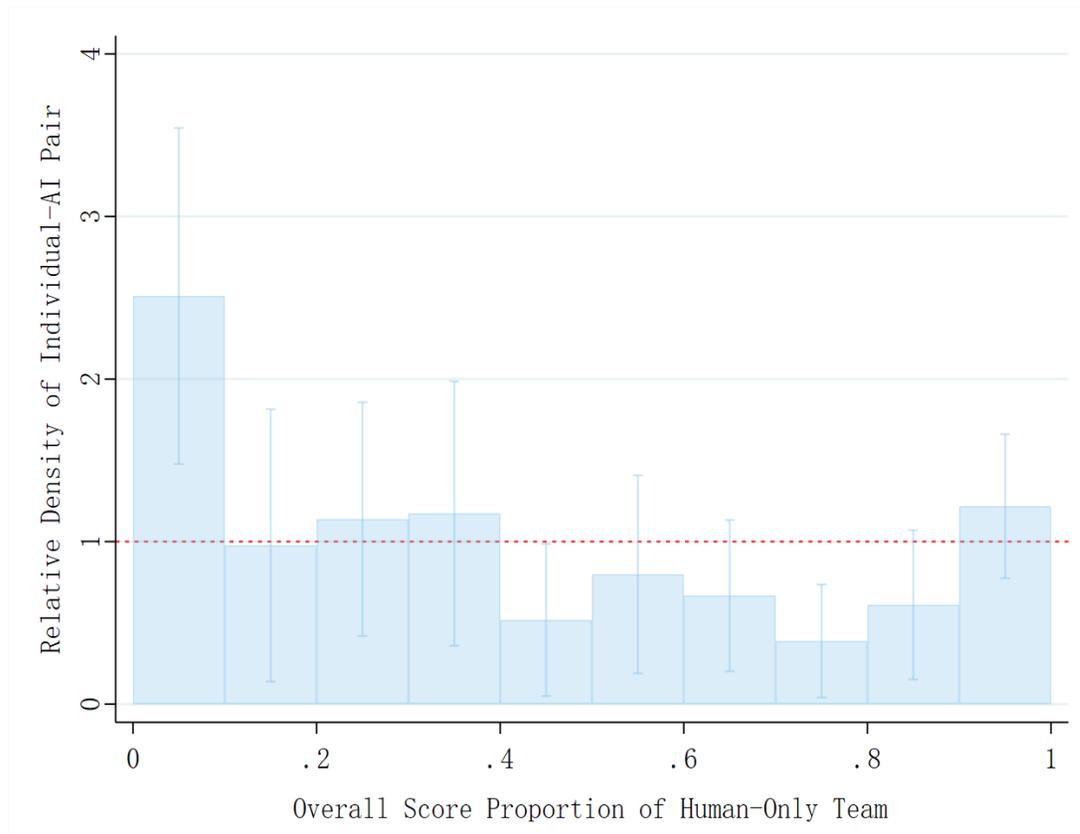
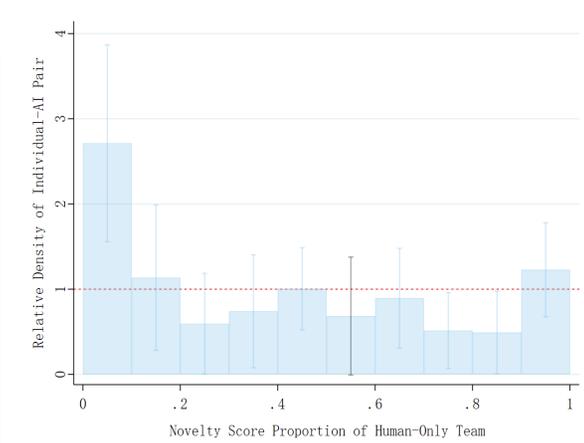
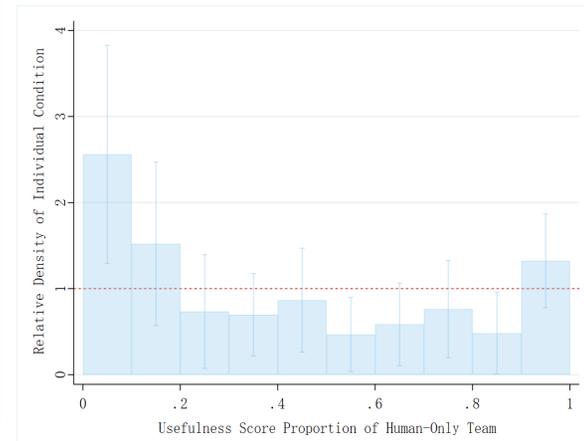

Note: The chart displays the relative distribution of individual-AI pairs and human-only teams on team performance, with the human-only teams as reference group. The left side shows the scores for the Overall dimension, while the right side, from top to bottom, displays scores for the Novelty and Usefulness dimensions. The red horizontal line at the value of 1 serves as a reference line. When the relative density value exceeds 1, it indicates a higher frequency of occurrences at that quantile for individual-AI pairs; conversely, values less than 1 indicate a lower frequency. The vertical lines represent the 95% confidence intervals for relative density values at specific score proportion intervals.

**Figure 2 Relative Distribution of Individual-AI Pair and AI Team (Single & Multiple) Conditions**

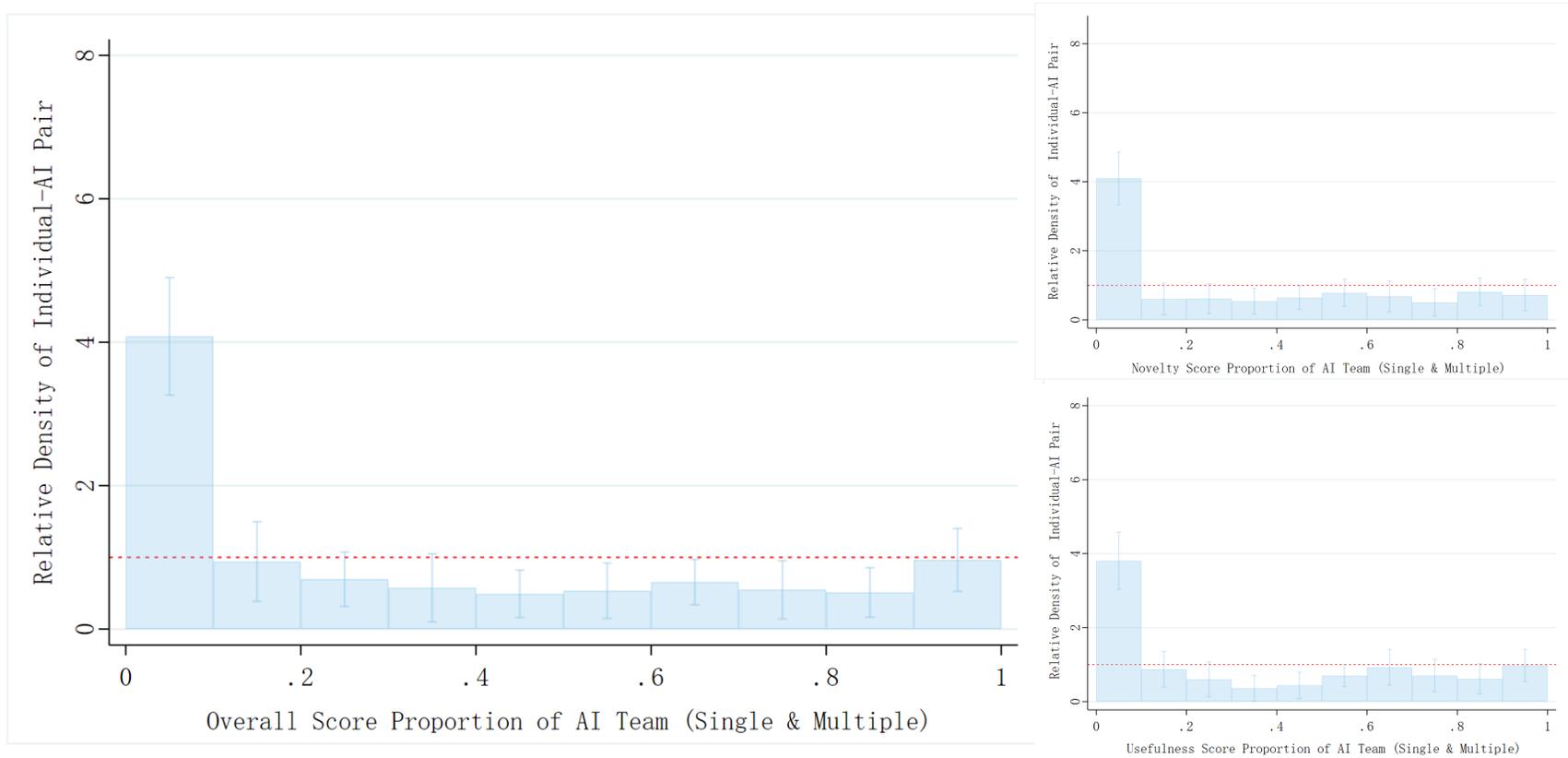

Note: The chart displays the relative distribution of individual-AI pairs and AI teams (single & multiple) on team performance, with the AI teams (single & multiple) as reference group. The left side shows the scores for the Overall dimension, while the right side, from top to bottom, displays scores for the Novelty and Usefulness dimensions. The red horizontal line at the value of 1 serves as a reference line. When the relative density value exceeds 1, it indicates a higher frequency of occurrences at that quantile for individual-AI pairs; conversely, values less than 1 indicate a lower frequency. The vertical lines represent the 95% confidence intervals for relative density values at specific score intervals.